\documentclass[preprintnumbers,aps,prd,twocolumn,superscriptaddress,notitlepage,nofootinbib]{revtex4-1}
\usepackage[english]{babel}
\usepackage[utf8]{inputenc}
\usepackage{amsthm}
\usepackage{mathtools}
\usepackage{physics}
\usepackage{xcolor}
\usepackage{graphicx}
\usepackage{adjustbox}
\usepackage{placeins}
\usepackage[T1]{fontenc}
\usepackage{lipsum}
\usepackage{csquotes}
\usepackage{physics} 
\usepackage[utf8]{inputenc}
\usepackage{amsmath}
\usepackage{slashed}

\newcommand{\nl}{\nonumber \\[5pt]}
\newcommand{\bn}{\bar{n}}

\newcommand{\bmat}[1]{\boldsymbol{#1}}

\newcommand{\lb}{\Big{\lbrack}}
\newcommand{\rb}{\Big{\rbrack}}
\newcommand{\lp}{\Big{(}}
\newcommand{\rp}{\Big{)}}
\newcommand{\lbc}{\Big{\lbrace}}
\newcommand{\rbc}{\Big{\rbrace}}

\newcommand{\Rangle}{\Big{\rangle}}
\newcommand{\Langle}{\Big{\langle}}
\newcommand{\eq}[1]{eq.~(\ref{eq:#1})}
\newcommand{\fig}[1]{fig.~\ref{fig:#1}}
\renewcommand{\sec}[1]{sec.~\ref{sec:#1}}

\newcommand{\zcut}{z_\text{cut}}
\newcommand{\Gcusp}{\gamma_{\text{cusp}}}

\newcommand{\Csoft}{\mathcal{C}}
\newcommand{\Usoft}{\mathcal{U}}

\newcommand{\LB}{\ln\lp\frac{\mu^2}{Q^2 \zcut} \rp}
\newcommand{\mgr}{m_{\text{gr.}}^2}

\begin{document}

%\title{Event based grooming and Winner-Take-All axis in DIS}
\title{Revisiting the role of grooming in DIS}
%%%%%%%%%%%%%%%%%%%%%%%%%%%%%%%%%%%%%%%%%%%%%%%%%%%%%%%%%%%%%%%%%%%%%%%%%%%%%%%%%%
\author{Y.~Makris}
\email{yiannis.makris@pv.infn.it}
\affiliation{INFN Sezione di Pavia, via Bassi 6, I-27100 Pavia, Italy}

%%%%%%%%%%%%%%%%%%%%%%%%%%%%%%%%%%%%%%%%%%%%%%%%%%%%%%%%%%%%%%%%%%%%%%%%%%%%%%%%%%
\begin{abstract}
    We introduce a novel grooming procedure, which is an extension of the  modified MassDrop tagging algorithm, tailored to the needs of deep inelastic scattering (DIS). The new algorithm, which grooms the event as a whole, takes advantage of the natural separation of current and target fragmentation in the Breit frame, in order to eliminate radiation in the beam and central rapidity regions.  We study the groomed invariant mass  in DIS and within soft-collinear effective theory we construct a  factorization theorem for the cross-section in the back-to-back limit. In this limit we show that, up to a normalization factor,  the cross-section does not depend on the incoming hadronic matrix element and we propose this measurement at HERA and the future electron-ion collider (EIC) as a probe to hadronization,  precision QCD, and cold nuclear matter effects. We also give an event based definition of the Winner-Take-All axis and comment on possible applications. 
\end{abstract}

\date{\today}

\maketitle
%%%%%%%%%%%%%%%%%%%%%%%%%%%%%%%%%%%%%%%%%%%%%%%%%%%%%%%%%%%%%%%%%%%%%%%%%%%%%%%%%%
\section{Introduction}
\FloatBarrier

Some of the main goals of the future Electron-Ion-Collider~\cite{Accardi:2012qut} are the study of hadronization, QCD dynamics, and 3D-imaging of the nucleon and nuclei. Among others, event shapes, jet, and jet substructure observables are proposed to achieve these goals.  These proposals are often motivated by the fact that event shapes and jet observables have already been applied successfully in various aspects of QCD phenomenology. However, the applicability and the necessity for such observables strongly depends on the experiment and the objective in consideration. For example, while for the clean environment of lepton-lepton colliders, event shape measurements can be directly compared to theoretical predictions, in hadronic colliders such comparison is spoiled by the contamination from underlying event (UE) and soft multi-parton interactions. A remedy to this problem is to use jet and jet substructure observables which are often extension of event shapes but limited only to the particles within the jets. Nonetheless, soft and uniformly distributed radiation can still significantly modify jet substructure measurements. To this end, theorists and experimentalists often employ grooming techniques~\cite{Krohn:2009th,Ellis:2009me,Dasgupta:2013ihk,Cacciari:2014gra,Larkoski:2014wba,Frye:2017yrw,Dreyer:2018tjj} to remove wide angle soft radiation, isolating this way the energetic core of the jet which is usually found near the jet axis. Groomed jets are shown to be mostly insensitive to the contamination from UE in hadronic colliders and thus constitute a robust framework for comparison between theory and experiment. In the past few years there has been a significant theoretical progress~\cite{Larkoski:2017cqq,Larkoski:2017iuy,Makris:2017arq,Baron:2018nfz,Kang:2018vgn,Makris:2018npl,Kardos:2018kth,Chay:2018pvp,Napoletano:2018ohv,Lee:2019lge,Hoang:2019ceu,Gutierrez-Reyes:2019msa,Marzani:2019evv,Mehtar-Tani:2019rrk,Kardos:2020ppl,Larkoski:2020wgx,Lifson:2020gua,Cal:2020flh} in our understanding of groomed observables. These studies were further  complemented by experimental results~\cite{Aaboud:2017qwh,Sirunyan:2018gct,Sirunyan:2018xdh,Aad:2019vyi,Aad:2020zcn,Adam:2020kug} both in $pp$ and $AA$ collisions. 

Groomed jet substructure observables have also been considered for the EIC. However, the role and the applicability of grooming at EIC is still a new and unexplored subject. While at EIC contamination from UE is not going to be an issue, there are many other motivations for the use of grooming techniques. Some of those are: i) constructing observables free from non-global-logarithms, ii) mitigation of hadronization corrections, iii) phenomenological handle on soft radiation, and  iv) as a dial for non-perturbative contributions. Despite the many motivations for the use of grooming, at EIC we anticipate jets with rather low particle multiplicities. In this case is unclear what  the effect of grooming will be and how  groomed jet observables should be interpreted.  In this paper we introduce an event-level grooming procedure which restores the role of grooming at EIC.  To achieve that we modify a widely used grooming algorithm, the  modified MassDrop tagging (mMDT)~\cite{Dasgupta:2013ihk}. By creating an angular ordering, measuring the angle from the direction of the photon in the Breit frame, the proposed grooming algorithm removes  radiation away from the struck quark direction and close to proton remnants. Then, groomed events at EIC can  be treated in an equivalent manner to groomed jets. 

Beyond the motivations mentioned above, the event-grooming algorithm we discuss in this paper, can be used for eliminating beam remnants and initial state radiation which  can be a significant source of uncertainty to global event observables at EIC and HERA.  As we discuss later using the example of groomed invariant mass, the shape of the groomed distribution is insensitive to the initial state hadronic matrix element and the kinematic variables of the hard process. Furthermore, the groomed distributions are also insensitive to the activity  in the forward region where (due to detector acceptance) constraints have been implemented at HERA and are expected to be implemented at the future electron-ion-collider (EIC) as well.

This paper is organized as follows. In \sec{alg} we briefly introduce our notation, review the DIS kinematics in the Breit frame, and we introduce the new grooming algorithm. Then we discuss the application of our algorithm to event shape observables using as an example the groomed invariant mass. We formulate a factorization theorem in the back-to-back limit within the soft and collinear effective theory (SCET)~\cite{Bauer:2000ew,Bauer:2000yr,Bauer:2001ct,Bauer:2001yt,Beneke:2002ph}, we compute the next-to- (NLL) and next-to-next-to-leading logarithmic (NNLL) accuracy distributions, and compare against the partonic shower of \textsc{Pyhtia} 8. We close this section with a brief discussion on hadronization effects. In \sec{WTA} we give an event based definition of the Winner-Take-All (WTA) axis and we discuss possible applications. We finally conclude in \sec{conclusions}. In the appendix we have collected fixed order results of various elements of the factorized cross-section, as well as the  leading order full-QCD cross-section. 

%%%%%%%%%%%%%%%%%%%%%%%%%%%%%%%%%%%%%%%%%%%%%%%%%%%%%%%%%%%%%%%%%%%%%%%%%%%%%%%%%%
\section{The grooming algorithm}\label{sec:alg}
\begin{figure}[t!]
  \centerline{\includegraphics[width = 0.45 \textwidth]{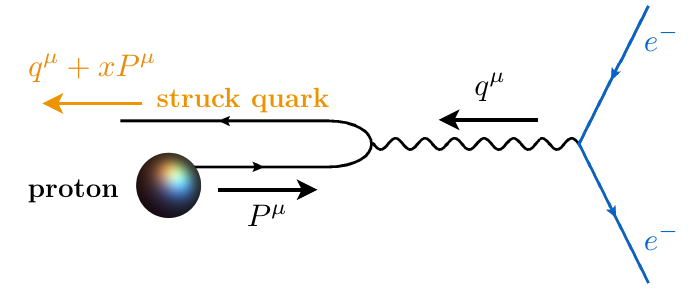}}
  \caption{~\label{fig:born-level} Diagrammatic illustration of the  Born level process in deep inelastic quark scattering in the Breit frame. }
\end{figure}

In this paper we work explicitly in the Breit frame, where exist a clean geometrical separation of target and current fragmentation.\footnote{However, all of the kinematic variables we introduce are either Lorentz invariant or boost invariant  along the direction of the proton in Breit frame. Thus, the grooming algorithm we develop will also be boost invariant along that same direction, e.g. in the $\gamma^*p$ center of mass frame.} To avoid contributions from the resolved photon events we will be considering only events with large photon virtuality, i.e., $Q = \sqrt{-q^2} \gg 1$ GeV, where $q^{\mu}$ is the four-momentum of the virtual photon. In the Breit frame we have,
\begin{equation}
    q^{\mu}  = \frac{Q}{2}( \bn^{\mu} - n^{\mu} ) = Q (0,0,0,-1)\,,
\end{equation}
where $n^{\mu} \equiv (1,0,0,+1)$ and $\bn^{\mu} \equiv (1,0,0,-1)$.
The proton momentum (up to mass corrections) is,
\begin{equation}
    P^{\mu} \simeq\frac{Q}{2x_B} n^{\mu} = \frac{Q}{2x_B} (1,0,0,+1)\,,
\end{equation} 
with $x_B$ the standard Bjorken variable, $x_B \equiv Q^2 / (2 \,q\cdot P)$. At Born level, the struck quark back-scatters against the proton and has momentum ($x\simeq x_B$):
\begin{equation}
    p_q^{\mu} = x P^{\mu} + q^{\mu} \simeq (Q/2) \bn^{\mu}\,.
\end{equation}
The fragmentation of the struck-quark leads to collimated (jet-like) radiation  pointing to the opposite of the beam direction as illustrated in \fig{born-level}. 
On the other hand, initial state radiation and beam remnants are moving in the opposite direction close to the proton’s direction of motion. It is this feature of the Breit frame, which leads to clean separation of target and current fragmentation, that we utilize in the new grooming algorithm.

Throughout this paper we will use the Lorentz invariant momentum fraction $z_i$,
\begin{equation}\label{eq:mom-var}
    z_i = \frac{P \cdot p_i}{P \cdot q} \;\;\;\xrightarrow[\text{frame}]{\text{Breit}} \;\;\; z_i = n\cdot p_{i}/Q = p^+_i / Q \,.
\end{equation}
where $p_i^{\mu}$ is the four-momentum of the particle $i$. Here and in the rest of this paper we use the standard notation, $p^{+} \equiv n \cdot p$ and $p^{-} \equiv \bar{n} \cdot p$. Note that from conservation of momentum we have 
\begin{equation}\label{eq:sum-rule}
    \sum_{i} z_i = 1\,,
\end{equation}
where the sum extents over all final state particles in the event, excluding the scattered lepton or its decay products. The struck quark fragments, found close the direction of the virtual photon share the largest fraction of the lightcone momenta $p^+$ and thus satisfy $z_i \sim 1$. Then, the constraint in \eq{sum-rule} requires that soft radiation and particles close to the direction of the target hadron are described by parametrically smaller values of the momentum fraction $z_i \ll 1$.

\begin{figure*}[t!]
  \centerline{\includegraphics[width = 0.95 \textwidth]{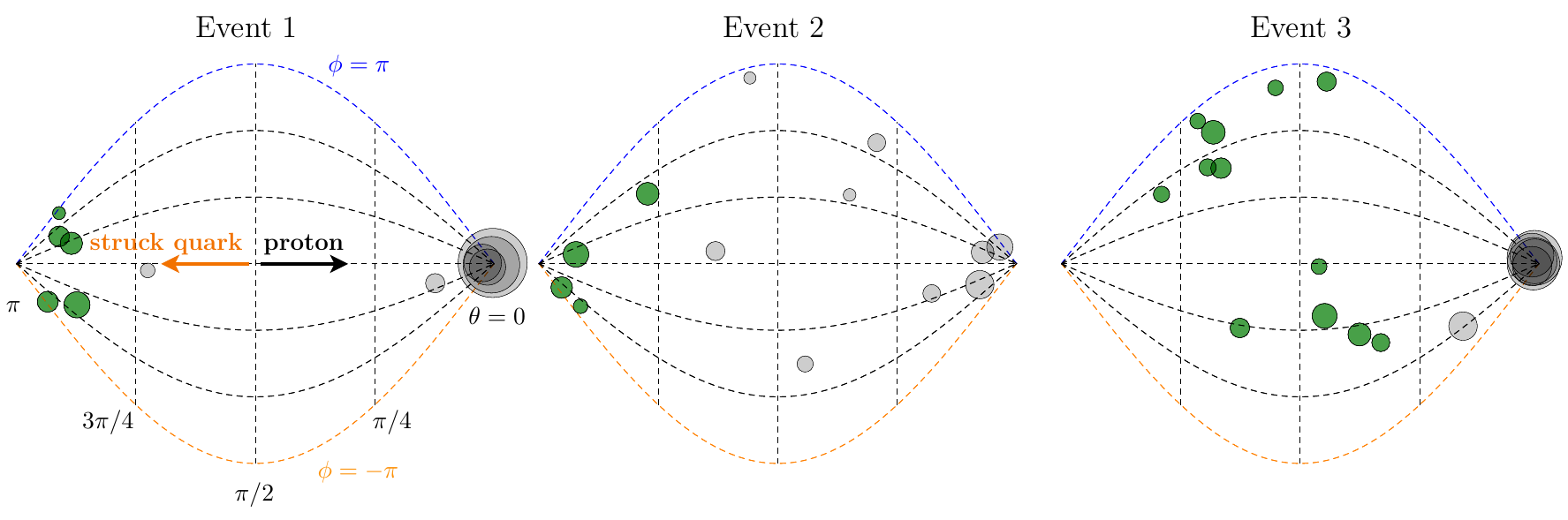}}
  \caption{~\label{fig:examples} Visualization of three \textsc{Pythia} 8 events at $\sqrt{s} = 63$ GeV and $Q\sim 10$ GeV before and after grooming. The particles in this events are represented by disks on the unfolded sphere. Green disks represent particles that pass grooming where grayed-out particles are removed from the event by the grooming procedure. For the grooming parameter we use here $\zcut = 0.1$}
\end{figure*}

%%%%%%%%%%%%%%%%%%%%%%%%%%%%%%%%%%%%%%%%%%%%%%%%%%%%%%%%%%%%%%%%%%%%%%%%%%%%%%%%%%

\subsection{Clustering and declustering}

We aim to remove radiation in the region close to the beam, but not in the opposite direction where the struck quark fragments are found. Therefore, the grooming procedure we introduce it is necessarily asymmetric under $\hat{z} \to -\hat{z}$.  Following a similar procedure to mMDT and SoftDrop, all particles in the event (excluding the scattered lepton or its decay products) are organized in a clustering tree based on their angular separation and their distance from  the direction of the virtual photon. To do exactly that we use the Centauro measure introduced in ref.~\cite{Arratia:2020ssx} in the context of jet algorithms for DIS in the Breit frame. The clustering procedure is:
\begin{enumerate}
    \item For every pair of particles $\{i,j\}$ in the event, we calculate the Centauro measure, 
    \begin{align}
       %\hspace*{1cm}  
       d_{ij} &= (\Delta \bar{\eta}_{ij})^2 +2 \bar{\eta}_i \bar{\eta}_j (1-\cos \Delta \phi_{ij})\;, 
    \end{align}
    where 
    \begin{equation}
        \bar{\eta}_i \equiv 2\sqrt{ 1+  \frac{q \cdot p_i}{ x_B P \cdot p_i} }\;\;\;\xrightarrow[\text{frame}]{\text{Breit}} \;\;\;\frac{2p_i^{\perp}}{p_i^+}\;,
    \end{equation}
    and $\Delta \bar{\eta}_{ij} = \bar{\eta}_i -\bar{\eta}_j$. Note that in the very backward limit (close to the photon axis) $\bar{\eta}_i$ is up to power-corrections the angle of the particle $i$ and the photon axis. 
    \item We find the minimum of all  $d_{ij}$ and we merge the particles $i$ and $j$ into a new ``branch'' by adding their momenta and removing the merged particles $i$ and $j$ from the list.
    \item Repeat until all particles in the event are merged together. 
\end{enumerate}
Keeping the history of all mergings, maps the event into an angular ordering tree where nearby particles are merged together in early stages. Then, clusters of particles near to the direction of the photon are combined  first. Last in the tree, are clustered particles near the target hadron beam.  With this angular ordering in hand we may now start the declustering process. The procedure is similar to SoftDrop, with some modifications. We open the tree back up in the reverse order of clustering. At each stage of the declustering, we have
two branches available, label them $i$ and $j$. We require:
\begin{equation}
\label{eq:condition}
    \frac{\min(z_i, z_j)}{z_i+z_j} > \zcut\,,
\end{equation}
where $\zcut$ is the grooming parameter, and $z_i$ is the momentum fraction variable of the branch $i$ as defined in \eq{mom-var}. If the two branches fail this requirement, the branch with smaller $z_i$ is removed from the event, and we decluster the other branch, once again testing \eq{condition}. The pruning continues until we have a branch that when declustered passes the condition in \eq{condition}.

We demonstrate the result of grooming in \fig{examples}, where we visualize a sample of events as simulated in \textsc{Pythia} 8 and groomed according to the procedure described above.  In this figure, each particle is illustrated as a disk, with area proportional to its energy, projected onto the unfolded sphere about the hard-scattering vertex. The vertical dashed lines correspond to constant $\theta$ and curved lines to constant $\phi$.  Grayed out disks correspond to particles that have been ``groomed away'' where green disks constitute the groomed event. 
Characteristic events are event 1 and 2, where the wide-angle particles are groomed away while the energetic and collinear particles (close to direction of the struck quark) pass grooming. However, for (relatively rare) events where an energetic cluster of particles exist in central rapidity regions (see for example event 3 in \fig{examples}), grooming will have small effect at central rapidities. This is due to the fact that such events contain central rapidity jets with relatively large momentum fraction $z_i$ and capable of passing the momentum-fraction condition in \eq{condition}.  The deferment event radiation patterns  can be  identified by considering measurements of event shapes which we discuss next.

%%%%%%%%%%%%%%%%%%%%%%%%%%%%%%%%%%%%%%%%%%%%%%%%%%%%%%%%%%%%%%%%%%%%%%%%%%%%%%%%%%
\subsection{Groomed invariant mass: A case study}

\begin{figure*}[t!]
  \centerline{\includegraphics[width =  \textwidth]{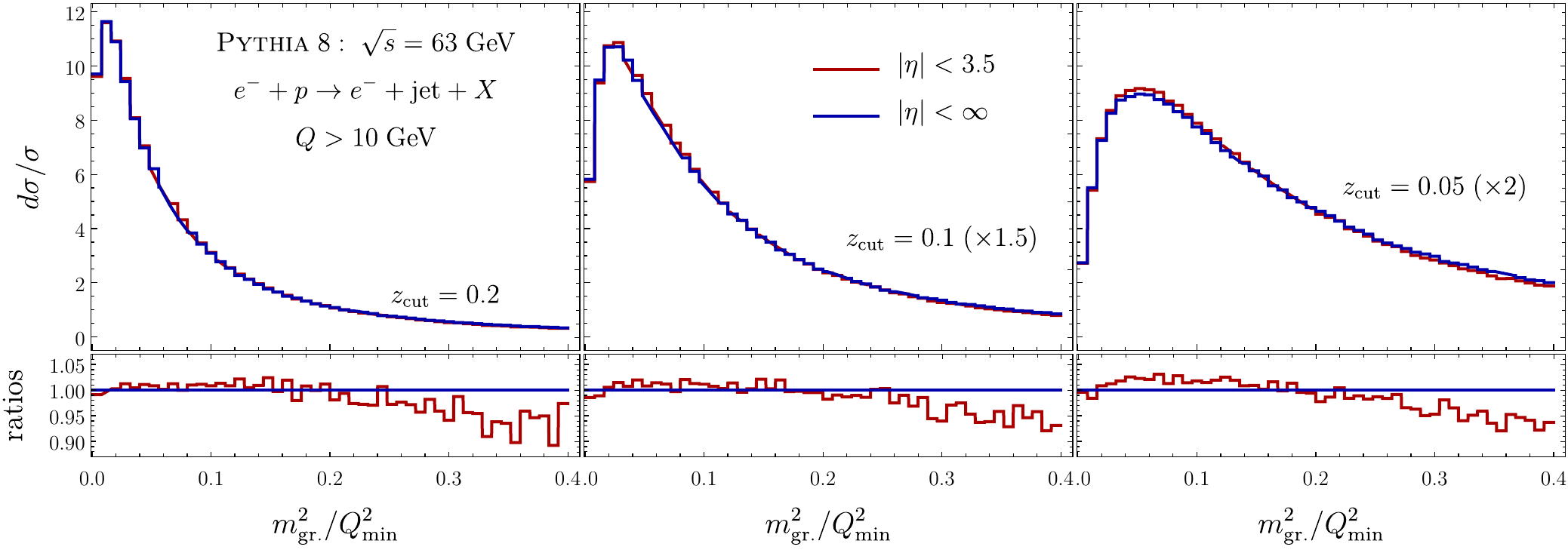}}
  \caption{~\label{fig:grVSun} \textsc{Pythia} 8 groomed invariant mass (GIM) spectrum for EIC kinematics. We present three different values of the grooming parameter $\zcut =$0.2, 0.1, and 0.05 from left to right. We compare the GIM distributions before and after imposing rapidity cutoff at $|\eta| < 3.5$ in the Laboratory frame.}
\end{figure*}

In the case of multiple  final-state hadrons, it is hard to study and understand the collective behaviour and correlations between those hadrons by simply looking at individual particle properties such as four-momenta and charge. It is therefore  useful to have a single variable, the value of which, characterizes the energy flow and distribution of momenta in space. Event shape variables play exactly this role. For the past several decades, there has been many event shape observables proposed and studied extensively, and they have played a central part in our understudying of Quantum Chromodynamics (QCD) with  plethora of applications in  QCD phenomenology. Some important examples of such applications are: i) determination of the QCD fundamental constants such as the strong coupling~\cite{Abreu:1999rc,MovillaFernandez:1997fr,Abbiendi:2004qz,Davison:2008vx,Abbate:2010xh,Abbate:2012jh,Gehrmann:2012sc,Hoang:2015hka} and the heavy quark masses~\cite {Akers:1994xi,Rodrigo:1997gy,Abreu:1997ey,Barate:2000ab,Abbiendi:2001tw}, ii) testing perturbative QCD~\cite {Schlatter:1981aw,Adeva:1983ur,Catani:1992jc,Abreu:1999rc,Abdallah:2003xz,Achard:2004sv,GehrmannDeRidder:2009dp,Weinzierl:2009ms}, iii) studying non-pertutbative effects and hadronization~\cite {Lee:2006fn,Lee:2006nr,Gehrmann:2009eh,Becher:2013iya}, and iv) tuning Monte-Carlo event generators~\cite{Akrawy:1990yx,Wraight:2011ej,Skands:2014pea,Bahr:2008pv}.

The role of event shape studies in this work is twofold. First, we are considering  groomed event shapes as a possible measurement in a future analysis  at HERA and at the EIC, and second as a theoretical tool for quantitatively investigating  the various features of the grooming algorithm that we introduced in this work. For our purposes we will focus on the groomed invariant mass (GIM) which we denote with $\mgr$ and is given by the invariant mass squared of the groomed event, 
\begin{equation}
   \mgr \equiv \lp \sum_{i} p_i^{\mu}\rp^2\;,
\end{equation}
where the sum runs over all particles that pass grooming, excluding the scattered lepton or its decay products. Note that if the sum included all hadrons in the event then this definition would correspond to the photon-proton center-of-mass energy, $W^2 \equiv (P^{\mu}+ q^{\mu})^2$, which can be uniquely determined from the kinematics of the hard process. However, GIM depends on the distribution of radiation in the event and is a measure of how well collimated is the radiation that passes grooming. For example, in the back-to-back limit where radiation is distributed in two pencil-like ``jets'' (from which one jet is the beam, see for example event 1 in \fig{examples}), we have $\mgr \ll Q^2$. On the other hand, in a dijet configuration (i.e., two final state jets, see for example event 3 in \fig{examples})  we have $\mgr \sim Q^2$.   We will study the spectrum of the GIM in different kinematic regimes and for a range of  values of the grooming parameter, $\zcut$. For these studies we are employing monte-carlo event generators as well as effective field theory (EFT) techniques.

We first consider \textsc{Pythia} 8 simulations  for potential EIC beam energies: 10 GeV electron on a 100 GeV proton ($\sqrt{s} \simeq 63$  GeV). After imposing the rapidity cuts on the final state particles in the Laboratory frame we perform the transformation in the Breit frame where we construct the observable and impose the grooming algorithm. We only focus on events with photon virtuality $Q > 10$ GeV and we test the effect of cutoff on the pseudo-rapidities, $|\eta| < \;3.5$. The results are shown in \fig{grVSun}. The GIM distributions are shown for three values of the grooming parameter $\zcut = 0.2$, 0.1, and 0.05 from left to right respectively. In each plot we show the results for $|\eta| < \;3.5$ (red) and no rapidity cuts (blue).  To improve readability, the distributions are normalized  and the horizontal axis has been re-scaled by the square of the minimum value of $Q$ included in the analysis, in this case  $Q_{\min}^2 = 100\;\text{GeV}^2$. It is very clear from these plots that the  distributions shift to smaller GIM values with larger $\zcut$. This is expected due to the additive property of the observable.  The correction to the distributions due to the rapidity cutoff  at $|\eta| < 3.5$ and for all values of $\zcut$ is found to be 1-2\% in the back-to-back limit and about 5\% in the tail of the distribution.\footnote{The effects are anticipated to be larger in the tail of the distribution since this region is populated by uniformly distributed radiation and dijet configurations which are more sensitive to radiation in the forward region.} This demonstrates that the groomed events are practically insensitive to rapidity cutoffs due to effective removal of initial state radiation and beam remnants in the forward region. This property of the grooming algorithm could be proven particularly useful in measurements of global observables such as thrust, N-jettiness, angularities, e.t.c., which could suffer from large systematic uncertainties due to detector acceptance in the forward region.   
%%%%%%%%%%%%%%%%%%%%%%%%%%%%%%%%%%%%%%%%%%%%%%%%%%%%%%%%%%%%%%%%%%%%%%%%%%%%%%%%%%

\subsection{Factorization}

One of the powerful aspects of grooming algorithms, such as mMDT and SoftDrop, is that the corresponding groomed jet mass distributions can be calculated in perturbation theory at very high precision. The mMDT jet mass in $e^+ e^-$ has been calculated at fixed order up to NNLO~\cite{Kardos:2016pic, DelDuca:2016ily, DelDuca:2016csb,Kardos:2018kth} and up to NNNLL~\cite{Kardos:2020gty} in resummation. 

Similarly the GIM distribution can be calculated perturbatively in the regime $\mgr \gg \Lambda_{\text{QCD}}$, by matching onto the parton distribution functions.\footnote{In principle small hadronization correction are also anticipated here.} The order $\mathcal{O}(\alpha_s)$ result of this matching procedure is given in the appendix.
However, in the case where $\mgr/Q^2 \ll 1$ and/or $\zcut \ll 1$, logarithms of these quantities can induce enhancements of higher order terms and potentially ruin the convergence of the perturbative expansion. To observe this enhancement it is sufficient to look at the LO result,
\begin{widetext}
\begin{equation}\label{eq:singular}
    \frac{d\sigma}{dx dQ^2 d\mgr} \Bigg{|}_{\substack{\mgr \ll Q^2 , \\[3pt] \zcut \ll 1\;\;\;}}  =  
    \frac{\sigma_0(Q,y)}{Q^2} \lp C^{(1)}(\zcut,\mgr) \frac{\alpha_s(\mu)}{4\pi}\, \Theta\lp\frac{1-x}{x} - \frac{\mgr}{Q^2} \rp\, \sum_{i=q,\bar{q}} Q_i^2\, f_{i/P}(x,\mu) + \mathcal{O}(\alpha_s^2) \rp\,,
\end{equation}
\end{widetext}
where $Q_i$ is the charge of the quark flavor $i$ and $f_{i/P}$ are the corresponding proton PDFs. The coefficient $C^{(1)}$ is extracted from expanding the full QCD cross section in the region  $\mgr \ll Q^2$ and $ \zcut \ll 1$ keeping only the leading terms. Doing so we find, 
\begin{equation}
    C^{(1)}(\mgr, \zcut) = - 4 C_F \frac{Q^2}{\mgr} \lp \frac{3}{4}+\ln\zcut \rp \,.
\end{equation}
Evaluating the coefficient for typical values of $\mgr/Q^2$ and $\zcut$ we find $C^{(1)}\sim50-200$. Comparing this to the expansion parameter, $\alpha_s(20\,\text{GeV})/(4\pi)\sim 0.01$, we find that the convergence of the perturbative expansion in this region of phase-space might be slow or even unstable and therefore potentially  resulting in large theoretical uncertainties. Thus, it is crucial that we perform the resummation of these logarithms to all orders in perturbation theory. To achieve this we factorize the cross-section into various matrix elements each depending on a single physical scale (or multiple scales of similar size). The renormalization group equations satisfied by these matrix elements will then allow us to establish a systematic framework for the resummation of large logarithms up to the desired logarithmic accuracy.

\begin{figure}[t!]
  \centerline{\includegraphics[width = 0.48 \textwidth]{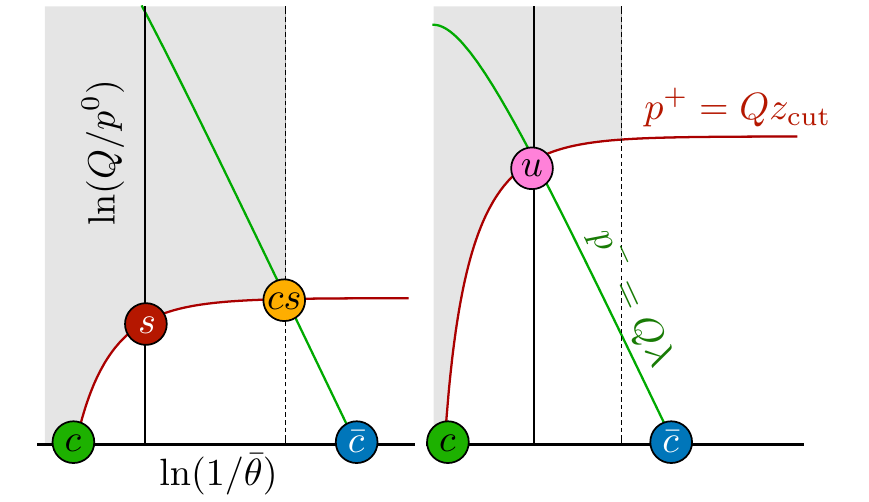}}
  \caption{~\label{fig:modes-A} The modes contributing to the factorization theorem of the groomed-invariant-mass spectrum in region 1 (top-panel) and region 2 (bottom panel). The modes are represented by locations on the energy and polar-angle plane, here $\bar{\theta} \equiv \pi -\theta$. The shaded area correspond to regions of phase-space failing grooming.}
\end{figure}

Here we will limit our analysis in the case where $\mgr \ll Q^2$ and $\zcut \ll 1$, and we construct the factorized cross-section for two possible hierarchies, 
\begin{align}\label{eq:regions}
    \text{region 1:}&\quad \quad 1 \gg \zcut  \gg \mgr/Q^2 \;,\nl
    \text{region 2:}&\quad \quad 1 \gg \zcut  \sim \,\mgr/Q^2 \;.
\end{align}
Although both regions are present in the GIM spectrum, region 1 is more relevant for larger values of $\zcut$ where region 2 better describes distributions with milder grooming. This is due to the fact that as we decrease $\zcut$, more radiation passes grooming, which then leads to larger values of $\mgr$. This can be observed in \fig{grVSun}.

%%%%%%%%%%%%%%%%%%%%%%%%%%%%%%%%%%%%%%%%%%%%%%%%%%%%%%%%%%%%%%%%%%%%%%%%%%%%%%%%%%
\subsubsection*{Region 1}

To establish the factorized form of the cross-section we work in soft collinear effective theory (SCET). We first identify  the degrees of freedom relevant to our observable in region 1 and then factorize the cross-section accordingly using the EFT techniques. We will not show the details of such a factorization procedure here but we will simply show the final result. Details for factorization within SCET in DIS can be found in refs.~\cite{Kang:2013nha, Gutierrez-Reyes:2019msa}. For the discussion that follows it is instructive to introduce a dimensionless variable, $\lambda$, 
\begin{equation}
    \lambda = \frac{\mgr}{Q^2}\;, 
\end{equation}
which will play the role of expansion and power-counting parameter of the EFT.

An important element of the grooming algorithm is that it organizes radiation based on the angular separation and distribution in space. It is then critical to think of the relevant modes the same way.  
The collinear radiation from the fragmentation of the struck quark is found close to the photon axis and therefore will merge into the clustering tree at the very early stages of the clustering process. Also energetic particles along this direction are anticipated to have large $p^+$ component, leading to momentum fractions $z_a\sim 1$ and as a result will pass the energy-fraction condition,  thus contributing to the GIM measurement. We refer to these modes as ``$\bar{n}$-collinear''. At relatively wider angles, but still close to the direction of the virtual photon we can find radiation  energetic just enough to pass the grooming procedure, i.e., $z_a \sim \zcut$  and thus can also contribute to the measurement of GIM. We refer to these modes as ``collinear-soft''~\cite{Bauer:2011uc}. In the other side of the event (close to the beam direction) we find initial state radiation and beam remnants from the incoming target-hadron. While this radiation is in principle very energetic, its directionality suggests that it will be merged to the clustering tree last and thus tested first against the energy momentum-fraction condition. With $z_a \ll 1$ it is most certain that particles from this region of phase-space will not pass grooming. We refer to these modes as ``$n$-collinear''. Finally, radiation in the mid-rapidity region can only be soft, otherwise will lead to large values of GIM. Therefore, soft radiation is constrained to fail  grooming, so it does not induces contributions to the measurement beyond the hierarchy described in region 1.  The directionality, the energy fraction (based on if they pass or fail grooming), and the GIM measurement, fix the $p^+$ and $p^-$ component of each mode. The $p^{\perp}$ component is then fixed by the on-shell condition: $(p^{\perp})^2 \sim p^+ p^-$. Thus, for the modes we discussed here  we have the following momenta scaling, 
\begin{align}
    n\text{-collinear:}&\quad p_n \sim Q(\zcut, 1, \sqrt{\zcut}) \;,\nl
    \text{soft:}&\quad p_{s} \sim Q\zcut(1, 1, 1) \;,\nl
    \text{collinear-soft:}&\quad p_{cs} \sim Q(\lambda, \zcut, \sqrt{\zcut\lambda}) \;,\nl
    \bn\text{-collinear:}&\quad p_{\bn} \sim Q(1, \lambda,  \sqrt{\lambda}) \;,
\end{align}
where we used the standard notation $p=(p^+, p^-, p^{\perp})$.
The modes are shown on the $\ln(Q/p^0) -\ln(1/\bar{\theta})$ plane in \fig{modes-A} (left panel), where $\bar{\theta} = \pi -\theta$.

\begin{figure*}[t!]
  \centerline{\includegraphics[width =  \textwidth]{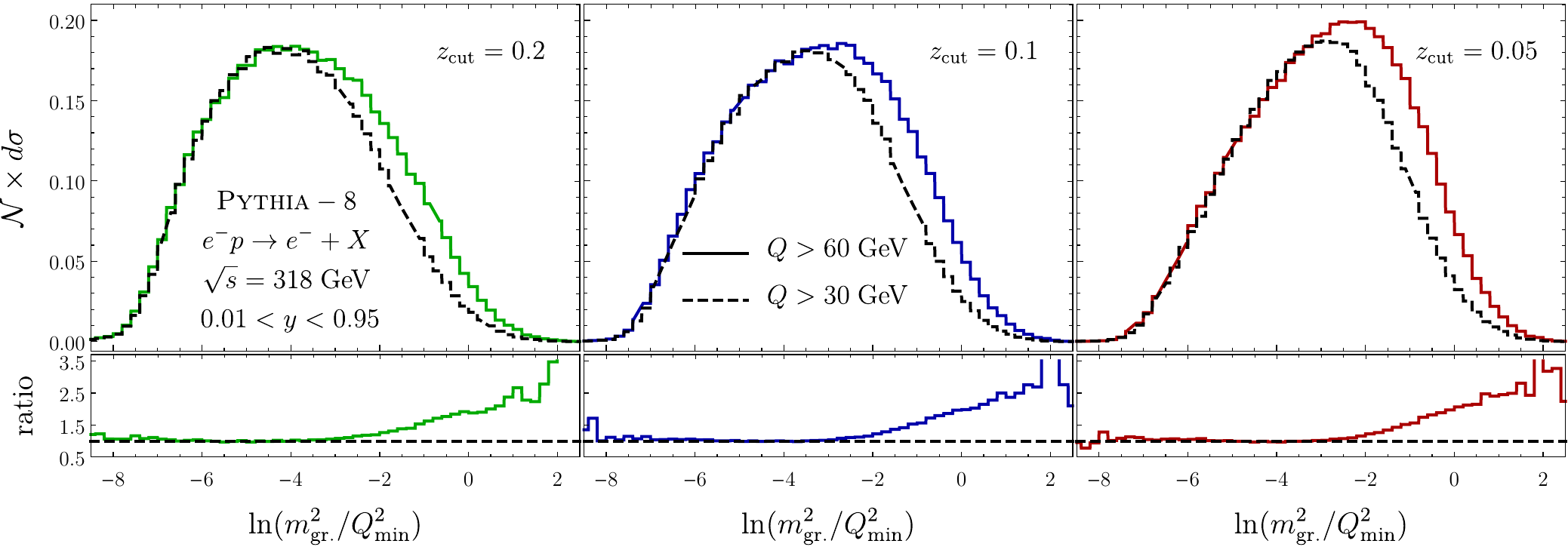}}
  \caption{~\label{fig:Q-dep} GIM for two values of $Q > $ 30 and 60 GeV for HERA beam energies: 27.5 GeV electron on 920 GeV proton. We find that for small values of the GIM spectrum the normalized distributions merge. This confirms what is anticipated from the form of the factorized cross-section as discussed in the main text.}
\end{figure*}

Considering these modes and following ref.~\cite{Kang:2013nha} we can factorize the cross-section into the EFT matrix elements and a hard function (which is derived by the matching of the EFT operators onto the QCD current). The factorized cross-section reads, 
\begin{widetext}
\begin{multline}\label{eq:x-sec}
    \frac{d\sigma}{dx dQ^2 d\mgr} =     H(Q,y,\mu) S(Q\zcut,\mu) \sum_{f}\mathcal{B}_{f/P}(x,Q^2\zcut,\mu)
     \int  de_{\bn} de_{cs} \; \delta(\mgr - e_{\bn} -e_{cs})\;  J(e_{\bn}, \mu^2 ) \;\Csoft ( e_{cs}\zcut,\mu^2 ) \\
     \times \lb 1 + \mathcal{O}\lp\zcut, \frac{\mgr}{\zcut Q^2}\rp \rb\;, 
\end{multline}
\end{widetext}
where $H$ is the hard function, $J$ is the quark thrust jet function, $\mathcal{C}$ is the SoftDrop collinear-soft function for jet-mass measurement,  $\mathcal{B}_{f/P}$ is the quark (flavor $f$) groomed beam function, and finally $S$ is the global soft function which up to clustering effects, in pure dimensional-regulator, is a scaleless function and thus its contribution to the factorization theorem starts at NNLL. 

The beam function~\cite{Fleming:2006cd} describes the  initial-state and $n$-collinear radiation constrained to be groomed away. For perturbative values of $Q\sqrt{\zcut}$ and up to power corrections, the beam function can be matched onto the collinear parton-distribution-functions (PDFs) ~\cite{Stewart:2009yx}, 
\begin{multline}\label{eq:matchingB-A}
    \mathcal{B}_{f/P}(x,Q^2\zcut,\mu) = \sum_i \int_{x}^{1} \frac{dz}{z} \;\mathcal{I}_{f/i}\lp \frac{x}{z},Q^2\zcut,\mu\rp \\
   \times f_{i/P}(z,\mu) \lb 1+\mathcal{O}\lp \frac{\Lambda_{\text{QCD}}^2}{Q^2 \zcut}\rp  \rb\;.
\end{multline}
The perturbative result  of the matching coefficients $\mathcal{I}_{f/i}$ up to NLO is the same as for the integrated beam-thrust measurement which we included in the appendix. Beyond NLO, where two or more final state partons are present, we need to consider clustering effects.  The sum over $f$ in \eq{x-sec} runs over all quark and antiquark flavours relevant to the scale we are working, where the sum over $i$ in \eq{matchingB-A} also includes the gluon.

Note that the cross-section in \eq{x-sec} is differential in $x$ and $Q^2$ as well as in $\mgr$.  In order to match  experimental measurements which are binned over various ranges of $x$ and $Q^2$ one needs to perform the integration numerically. Here we propose an alternative approach that allow us to compare with experimental measurements without the need for integration in $x$ and $Q^2$ by simply looking at the normalized cross-section instead. This relies on the observation that the cross-section dependence on $x$ and $Q^2$ is isolated in the hard, beam, and soft functions. These functions do not depend on the GIM measurement. Therefore, the shape of the GIM distribution in region 1 is determined by the jet and collinear-soft function which are independent $x$ and $Q^2$. This is tested using \textsc{Pythia} simulations as shown in \fig{Q-dep}.

\begin{widetext}
The normalized cross-section is then defined as follows,
\begin{equation}
    \frac{d\sigma^{\text{norm.}}}{d\mgr} (\mgr,e_{\min},e_{\max}) \equiv  \lb \int_{e_{\min}}^{e_{\max}} d\mgr   \int_{x,Q^2} \frac{d\sigma}{dx dQ^2 d\mgr}   \rb^{-1} \times
    \int_{x,Q^2} \frac{d\sigma}{dx dQ^2 d\mgr} \;,
\end{equation}
where the integration over $x$ and $Q^2$ is determined based on the experimental  choice of  binning. Expanding in terms of the factorized cross-section we find 
\begin{equation}\label{eq:norm-x-sec}
    \frac{d\sigma^{\text{norm.}}}{d\mgr} (\mgr,e_{\min},e_{\max}) =\mathcal{N}(e_{\min},e_{\max},\mu) \int  de_{\bn} de_{cs} \; \delta(\mgr - e_{\bn} -e_{cs})\;  J(e_{\bn}, \mu^2 ) \;\Csoft ( e_{cs}\zcut,\mu^2 )\;,
\end{equation}
where 
\begin{equation}\label{eq:norm-final}
    \mathcal{N}^{-1}(e_{\min},e_{\max},\mu) = \int_{e_{\min}}^{e_{\max}} d\mgr  \int  de_{\bn} de_{cs} \; \delta(\mgr - e_{\bn} -e_{cs})\;  J(e_{\bn}, \mu^2 ) \;\Csoft ( e_{cs}\zcut,\mu^2 )\;.
\end{equation}
\end{widetext}
The moralized cross-section can now be calculated  without considering the integration over the variables $x$ and $Q^2$. There are some obvious practical advantages adapting this approach. First,  the calculation of the normalized cross-section can be achieved significantly easier and avoids systematic uncertainties  associated with PDF errors. Beyond these, considering the normalized cross section can help reduce theoretical uncertainties since it allow us to push the perturbative calculation at the same order as  has been achieved for mMDT jet invariant mass (i.e., NNNLL). The latter statement relies on the fact that the Centauro measure in the collinear  limit (along the photon's direction) is, up to power corrections, equal to the C/A(SI) measure.

We point out that, although the expression in \eq{norm-final} does not explicitly depend on $Q$ the factorization only holds for $\mgr \ll Q^2 \zcut $ and thus the region of $\mgr$ for which we can use this result depends on the lowest value $Q$ which we are considering for a particular analysis. Therefore, we require that we normalize only within that same region, $\mgr < e_{\max} \ll Q_{\min}^{2} \zcut$, or within the region where power corrections to the factorization theorem remain small. 

%%%%%%%%%%%%%%%%%%%%%%%%%%%%%%%%%%%%%%%%%%%%%%%%%%%%%%%%%%%%%%%%%%%%%%%%%%%%%%%%%%
\subsubsection*{Region 2}

Going beyond region 1 the shape of the cross-section depends on both $Q^2$ and $x$. For example in region 2, which is described by the hierarchy of scales: $1\gg \zcut \sim \lambda$, the soft-collinear and soft radiation merge into a single mode which we will refer to us ultra-soft  (u-soft). The scaling of the momenta of the relevant modes is,
\begin{align}
    n\text{-collinear:}&\quad p_n \sim Q(\zcut, 1, \sqrt{\zcut}) \;,\nl
    \text{ultra-soft:}&\quad p_{us} \sim Q\zcut(1, 1, 1) \sim Q\lambda(1, 1, 1)\;,\nl
    \bn\text{-collinear:}&\quad p_{\bn} \sim Q(1, \lambda,  \sqrt{\lambda}) \;.
\end{align}
\begin{widetext}
These modes are shown on the $\ln(Q/p^0) -\ln(1/\bar{\theta})$ plane in \fig{modes-A} (right panel). In addition the veto on mid-rapidity soft radiation that pass grooming has to be lifted since u-soft modes may now contribute to the measurement. The resulting factorization theorem is,
\begin{multline}\label{eq:x-sec-2}
    \frac{d\sigma}{dx dQ^2 d\mgr} =     H(Q,y,\mu) \sum_{f}\mathcal{B}_{f/P}(x,Q^2\zcut,\mu)
     \int  de_{\bn} de_{us} \; \delta(\mgr - e_{\bn} -e_{us})\;  J(e_{\bn}, \mu^2 ) \;\Usoft ( e_{us}, Q, \zcut,\mu^2 ) \\
     \times \lb 1 + \mathcal{O}\lp\zcut, \frac{\mgr}{Q^2}\rp \rb \;.
\end{multline}
In the above equation the hard, beam, and jet functions are the same as in region 1, but the product of soft and collinear-soft functions has been replaced by the u-soft function, $\Usoft$, that now depends explicitly on the hard scale $Q$, the grooming parameter, $\zcut$, and the GIM measurement. At fixed order in the strong coupling expansion one can relate the two expressions within the hierarchy of region 1 up to power corrections,
\begin{equation}\label{eq:matchingS-CS}
    \Usoft(\mgr ,Q,\zcut,\mu^2) \Big{|}_{\mgr /Q^2 \ll \zcut} = S(Q \zcut,\mu) \Csoft(\mgr \zcut ,\mu^2) \Big{[}1+ \mathcal{O} \Big{(} \frac{\mgr}{Q^2 \zcut} \Big{)} \Big{]}\;.
\end{equation}
\end{widetext}
We evaluated the u-soft function at NLO and showed that the power corrections vanish at this order. Therefore we have that at NLO the u-soft function is simply given by the collinear-soft function of region 1,
\begin{equation}\label{eq:usoft-NLO}
    \Usoft(\mgr ,Q,\zcut,\mu^2) \Big{|}_{\text{NLO}} =  \Csoft(\mgr \zcut ,\mu^2)  \Big{|}_{\text{NLO}}\;.
\end{equation}
With this, we now have all the necessary ingredients to perform the resummation of all large logarithms in region 1 and region 2 up to NLL. 

%%%%%%%%%%%%%%%%%%%%%%%%%%%%%%%%%%%%%%%%%%%%%%%%%%%%%%%%%%%%%%%%%%%%%%%%%%%%%%%%%%
\subsubsection*{Power corrections}

\begin{figure}[t!]
  \centerline{\includegraphics[width = 0.48  \textwidth]{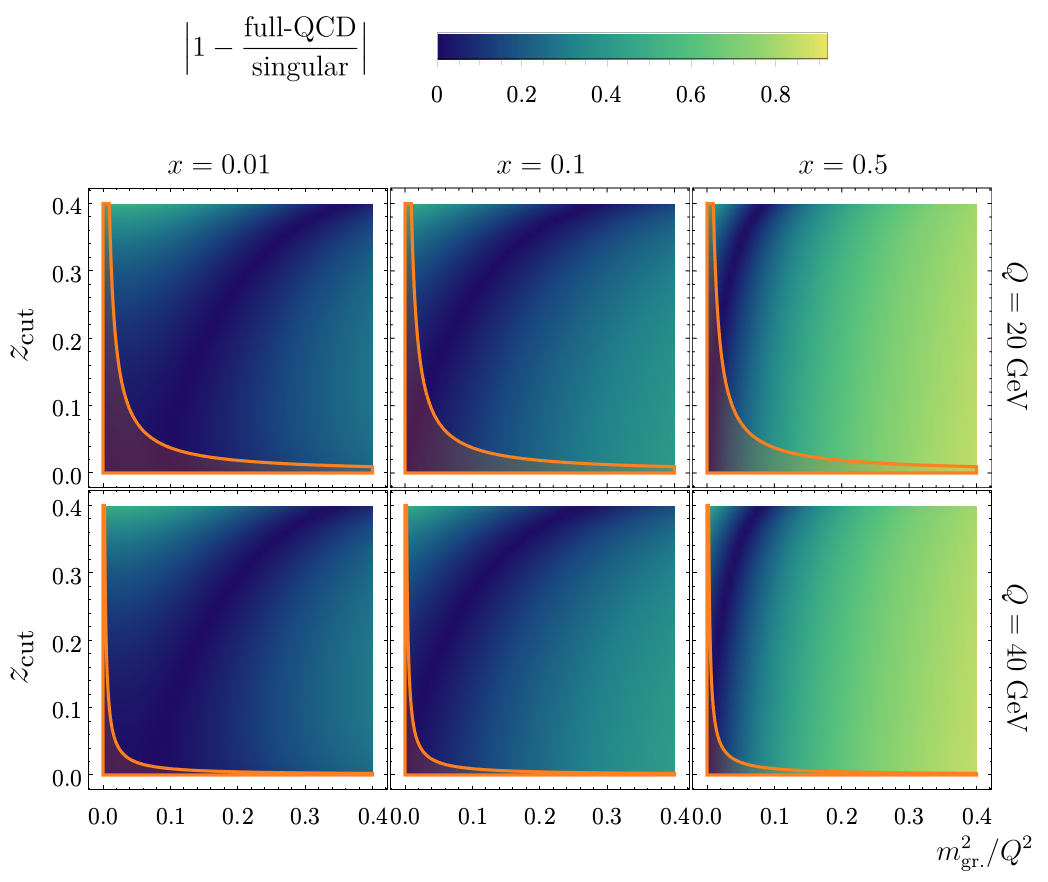}}
  \caption{~\label{fig:power-corr.} The relative size of power corrections as a function of $\mgr$ and $\zcut$ for different values of $x\text{-}Q^2$. The orange shaded area denotes the non-perturbative region for which $\mgr \zcut \leq 1.5$ GeV$^2$. }
\end{figure}

Before we start considering numerical results is important to discuss the role of power corrections to the factorization theorems we analyzed in this section. These corrections rise due to the fact that we have effectively expanded the QCD amplitudes in particular kinematic regimes, specifically the ones shown in \eq{regions}. One can estimate the size of these corrections order-by-order in perturbation theory by subtracting the singular limit in \eq{singular} from the full QCD result in \eq{fullQCD}. For different values of $x$ and $Q^2$ and as a function of $\zcut$ and $\mgr/Q^2$  we plot the relative corrections in \fig{power-corr.}. We find that for  $x \lesssim 0.1$ the power corrections decrease with smaller $x$ and remain small over a wide range of $\zcut$ and $\mgr/Q^2$. However, the corrections seem to increase significantly in the region $x\sim 1$. This increase is due to the fact that at large $x$ and at large invariant mass the cross section is suppressed by the constraint,
\begin{equation}
    \mgr \leq W^2 \simeq \frac{1-x}{x} Q^2\,.
\end{equation}
The non-perturbative regime is denoted with the orange shading in \fig{power-corr.}  which corresponds to the most IR scale of the problem to be of order 1 GeV  (in this case $\mu_{\text{IR}}^2 \sim \mgr \zcut \sim 1$ GeV).

Such corrections can in principle be included in the resummed result order-by-order in the strong coupling expansion, if the full QCD result is known either analytically or numerically. This can be done in a matching procedure where we add the fixed order corrections to the resummed spectrum after turning off evolution in the region $\mgr \gtrsim Q^2$. This way one can obtain a reasonable prediction in the whole range of the GIM spectrum. Note however that, in-contrast to the standard groomed observables,\footnote{The standard groomed jet-mass observable does not depend on the grooming algorithm at sufficiently large values of the jet-mass, creating a cusp in the distribution where the transition occurs. The same we would have observed if we studied the groomed 1-jettiness for example instead of the GIM.} the observable we discuss here depends on the grooming parameters in the whole GIM spectrum, $0<\mgr/Q^2 < (1-x)/x$. Thus, logarithmic enhancements for $\zcut \ll 1$ remain relevant through-out the whole spectrum and as a result resummation of these logarithms is important for even large values of GIM. For this reason in this work we will focus in the resummation regions 1 and 2 and we will not pursue the task of matching at the tail of the distribution.

%%%%%%%%%%%%%%%%%%%%%%%%%%%%%%%%%%%%%%%%%%%%%%%%%%%%%%%%%%%%%%%%%%%%%%%%%%%%%%%%%%
\subsection{Resummed results}

\begin{figure*}[t!]
  \centerline{\includegraphics[width =  \textwidth]{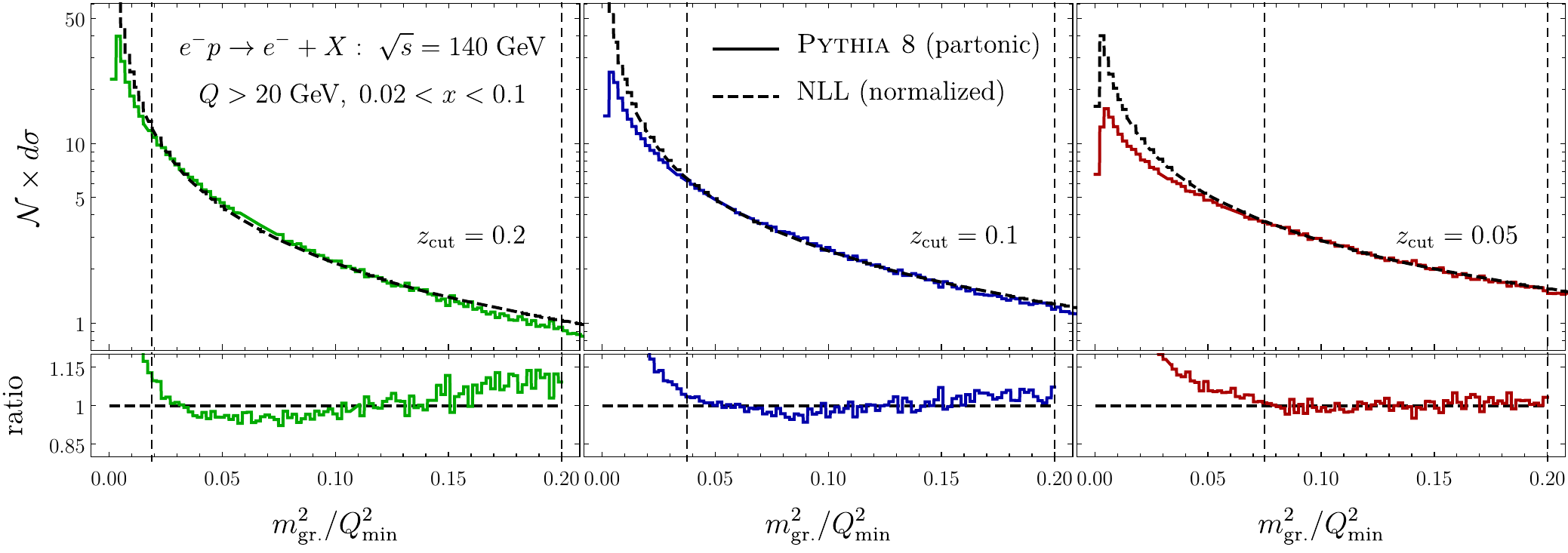}}
  \caption{~\label{fig:PythiaVSnll} Comparing the NLL predictions (dashed) against the partonic shower \textsc{Pythia} 8 (solid) for three different values of the grooming parameter $\zcut=$ 0.2, 0.1, and 0.05.}
\end{figure*}

\begin{figure*}[t!]
  \centerline{\includegraphics[width =  \textwidth]{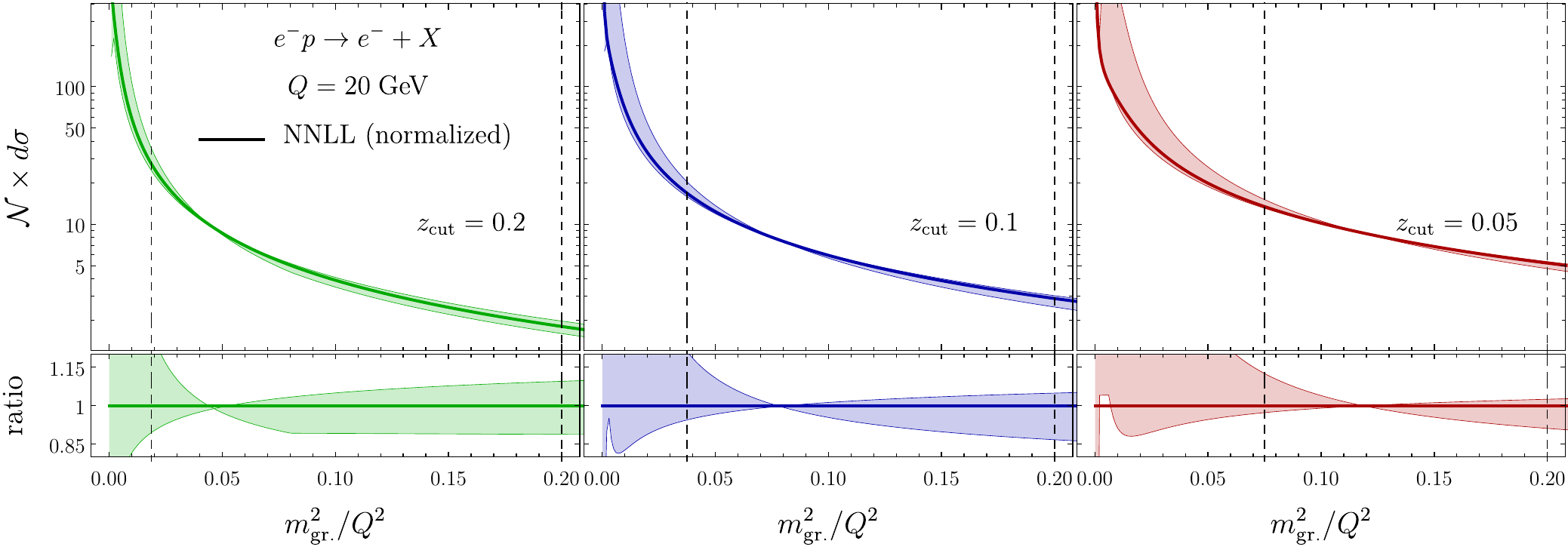}}
  \caption{~\label{fig:nnll} NNLL prediction including theoretical uncertainties. The theory band is constructed by varying all scales in the factorization theorem by a factor 2 and 1/2.}
\end{figure*}

The various elements of the factorization have renormalization scale, $\mu$, dependence. The resummed cross-section, where all large logarithms are resummed up to a particular logarithmic accuracy, is  obtained by evaluating the elements of factorization at their canonical scales and then use renormalization group (RG) equations to evolve  them up to a common scale. The RG equation satisfied by each of the relevant functions is, 
\begin{equation}
    \frac{d}{d\ln\mu}G(\mu) = \gamma_G(\mu) G(\mu)\;,
\end{equation}
where $G$ can be any of the functions that appear in the factorization theorems in \eq{x-sec} and \eq{x-sec-2}.\footnote{For functions that depend on the GIM measurement (i.e., $J$, $\Csoft$, and $\Usoft$) the above equation applies to the Laplace transformed expressions (w.r.t. the variable $e$), otherwise  the \emph{l.h.s.} can be  expressed as a convolution between $\gamma_G$ and $G$.} The anomalous dimensions  $\gamma_G(\mu)$ are usually written in terms of the cusp anomalous dimension, $\Gcusp[\alpha_s]$, and a non-cusp term, $\gamma_G[\alpha_s]$,
\begin{equation}
    \gamma_G(\mu) = d_G \Gcusp[\alpha_s] \ln \lp \frac{\mu}{m_G} \rp + \gamma_G[\alpha_s]\;,
\end{equation}
where $d_G$ is a number, $m_G$ has dimensions of mass and both $d_G$ and $m_G$ depend on the function $G$. The solution of the RG equation in Laplace space can be written as a product of the evolution kernel $U_G$ and the function $G$, evaluated at some initial scale $\mu_0$, 
\begin{align}
    G(\mu) &= G(\mu_0) \times \exp \lp \int_{\mu_0}^{\mu} d\ln \mu'\, \gamma_G(\mu') \rp \\
    &\equiv G(\mu_0) \times  U_G(\mu,\mu_0)\;.
\end{align}
The kernel $U_G(\mu,\mu_0)$ is computed at NLL and NNLL accuracy in terms of $d_G$, $m_G$, $\Gcusp[\alpha_s]$ and $\gamma_G[\alpha_s]$ in multiple references in literature (see for example refs.~\cite{Abbate:2010xh,Ligeti:2008ac,Frye:2016aiz}). The initial scale $\mu_0$ for each function is usually chosen to be the canonical scale, i.e., the scale that minimizes the logarithms in the fixed order expansion. For the functions we are considering here, the canonical scales are:
\begin{align}
    \mu_H = Q, \quad  \quad \mu_{\mathcal{B}} &= Q\sqrt{\zcut}, \quad  \quad \mu_S =Q \zcut \nl 
    \mu_J^2 = \mgr,  \quad & \quad \mu_{\Csoft}^2= \mu_{\Usoft}^2 = \mgr  \zcut\,.
\end{align}
For small values of the invariant mass the jet and collinear-soft scales become non-perturbative and the perturbative expansion breaks. For this reason, at small invariant masses, we freeze these scales before reaching the Landau pole using the following prescription:
\begin{equation}
    \mu_{G} \to g(\mu_G , a, b) = \frac{(1+(a \,\mu_G)^{b})^{1/b}}{a}\;,
\end{equation}
where we chose $a=2\;\text{GeV}^{-1}$ and $b=5$. 

Regarding the factorization in region 1, we collect from the literature all the necessary ingredients  for the construction of the NLL cross-section in the appendix. To achieve this accuracy we need the $\mathcal{O}(\alpha_s^2)$ cusp terms and the $\mathcal{O}(\alpha_s)$ non-cusp terms of the anomalous dimensions. For the NNLL predictions we need the $\mathcal{O}(\alpha_s^2)$ non-cusp anomalous dimensions, which we also have with the exception of the soft and beam functions. However,the normalized cross-section can still be calculated at NNLL with what is already known, since it is independent of the soft and beam functions. For region 2 we only have ingredients sufficient for a NLL calculation which in this case is identical to region 1. We compare our NLL  numerical results (region 1 and 2) for the normalized distributions and compare against the partonic shower of \textsc{Pythia} 8 in \fig{PythiaVSnll}. We give the NNLL normalized distributions (region 1) including theoretical uncertainties in \fig{nnll}.

We first consider the comparison  of NLL cross-section to \textsc{Pythia}'s partonic shower. This serves a consistency check of our formalism and we discuss hadronization effects later in this section. To make meaningful comparison of our perturbative result with \textsc{Pythia}, we require relatively large values of $Q$ where a perturbative regime exist. Particularly we choose $Q_{\min} = 20$ GeV and for this reason we consider the potential EIC kinematics with beam energies: 18 GeV electron on 275 GeV proton ($\sqrt{s} = 140.7$ GeV).  For the normalization of the cross-section the lower boundary  is fixed such that the softest scale of the problem (that is the collinear-soft scale, $\mu_{\Csoft}$)  remains within the perturbative regime. This leads to the constraint:   
\begin{equation} 
    e_{\min} =  \frac{\Lambda_{\text{NP}}^2}{\zcut}\;,
\end{equation}
where $\Lambda_{\text{NP}}$ is a non-perturbative scale of order $\gtrsim 1$ GeV. In this paper we chose $\Lambda_{\text{NP}}^2 = 1.5\;\text{GeV}^2$. The upper boundary  should in general be chosen such that power corrections to the factorization theorem in \eq{x-sec} remain small. For the purpose of this comparison we can extend the upper limit as long as we remain well within region 1 and 2, $\mgr \ll Q^{2}_{\min}$. Guided by the size of power corrections in \fig{power-corr.} here we chose,  
\begin{equation}
e_{\max} = 0.2  \;.
\end{equation}
For three different values of the grooming parameter $\zcut = 0.2$, 0.1, and 0.05 the comparison with \textsc{Pythia} is shown in \fig{PythiaVSnll}. The values $e_{\max}$ and $e_{\min}$ are shown with dashed vertical lines. We find excellent agreement of our NLL resummed result with the simulation. However, we see that the perturbative regime shrinks as we decrease $\zcut$. This effect will be exaggerated if we consider smaller values of $Q$ where the non-perturbative boundary, $e_{\min}$, moves to the right towards larger $\mgr/Q^2_{\min}$ values, further shrinking this way the perturbative regime. This should motivate further studies in both directions: in the perturbative tail of the distribution, to the right of $e_{\max}$, and in the non-perturbative regime, to the left of $e_{\min}$ where $\mgr \zcut \sim \Lambda^2_{\text{NP}}$.

\begin{figure*}[t!]
  \centerline{\includegraphics[width =  \textwidth]{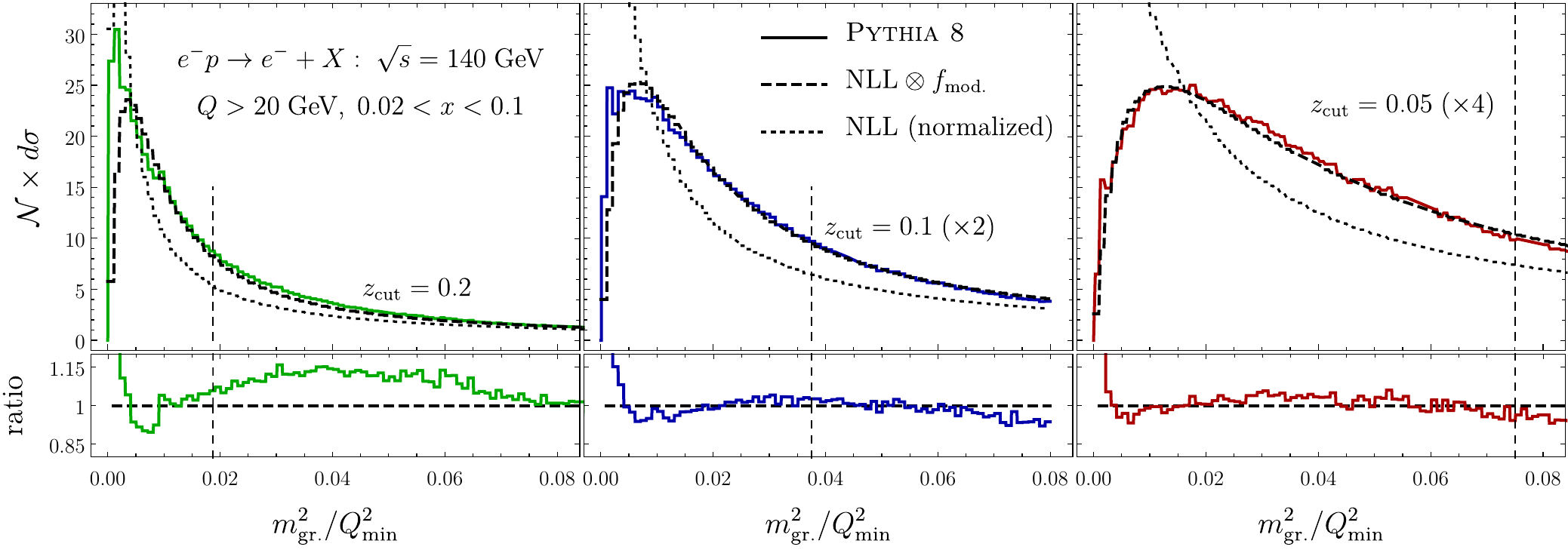}}
  \caption{~\label{fig:hadronization} Hadronic  \textsc{Pythia} simulations for GIM distribution compared against NLL convolved with shape function $f_{\text{mod.}}$  for EIC ($\sqrt{s} = 140$  GeV) kinematics. To emphasize the role of the shape function we also included the NLL distribution without including the model for hadronization.}
\end{figure*}

In \fig{nnll} we show the NNLL normalized  distribution in region 1. We have also estimated the theoretical uncertainty by multiplying the three scales $\mu_J$, $\mu_{\mathcal{C}}$, and $\mu$ by a factor of 2 and 1/2 separately. The largest deviations from the central curve are collected into the envelope  that represents the theoretical uncertainty. We find that within the perturbative regime the uncertainty varies between 10-15\%. Note that the size of this error bands is sensitive to the value of the non-perturbative scale $\Lambda_{\text{NP}}$ which sets the lower limit of integration. As this value decreases we incorporate more of the non-perturbative regime into the normalization factor resulting in large theoretical uncertainties. This is also apparent from the fact that the theoretical uncertainty below the $e_{\min}$ increases rapidly as shown in \fig{nnll}. 
%%%%%%%%%%%%%%%%%%%%%%%%%%%%%%%%%%%%%%%%%%%%%%%%%%%%%%%%%%%%%%%%%%%%%%%%%%%%%%%%%%
\subsection{Hadronization and non-perturbative corrections}

In the previous section, as a consistency test of our factorization we compared our resummed predictions against \textsc{Pythia} 8 partonic shower. In region 1 ($1\gg \zcut \gg \mgr/Q^2$), comparing our resummed distributions against \textsc{Pythia}'s hadronic spectrum we find significant deviation.  In this section we perform a proof-of-concept exercise where we incorporate hadronization effects through a non-perturbative shape function. 

Equipped with a factorization theorem and precise predictions of the perturbative spectrum, one can approach hadronization effects in a systematic and rigorous manner. A field-theoretic description and leading power approach to hadronization effects in groomed jet mass is given in ref.~\cite{Hoang:2019ceu}. The formalism is also applicable here since the collinear-soft function that appears in this work, and which dominates the non-perturbative contributions, is the same as in the SoftDrop ($\beta =0$) groomed jet-mass. In the formalism of ref.~\cite{Hoang:2019ceu}, two regions of the invariant mass are identified, 
\begin{align}
    \text{SDNP:}& \quad \mgr \zcut \lesssim \Lambda_{\text{QCD}}^2 \;,\nl
    \text{SDOE:}&\quad  \mgr \zcut \gg \Lambda_{\text{QCD}}^2\;.
\end{align}
While in the region SDNP hadronization corrections are encoded into a universal shape function, in the perturbative region SDOE the hadronization corrections are described by two non-perturbative matrix elements associated with the ``shift'' and ``boundary'' effects. In our  study for simplicity we focus only on the non-perturbative region SDNP which includes the peak of the distribution and where the hadronization effects are more apparent. Particularly  we incorporate a shape function for describing the hadronization corrections on the GIM spectrum and we use the parameterazation discussed in ref.~\cite{Frye:2016aiz},
\begin{equation}\label{eq:shape_convolution}
    \frac{d\sigma_{\text{had.}}}{dx dQ^2 d\mgr} = \int d\epsilon   \frac{d\sigma}{dx dQ^2 d\mgr} \lp \mgr - \frac{\epsilon^2}{ \zcut} \rp \, f_{\text{mod.}} (\epsilon)\;,
\end{equation}
where 
\begin{equation}
f_{\text{mod.}}(\epsilon) = N_{\text{mod.}} \frac{4 \epsilon}{\Omega^2} \exp\lp \frac{2 \epsilon}{\Omega}\rp\;,
\end{equation}
is the model shape function which depends on two non-perturbative parameters: the mean of the shape function $\Omega$ and the normalization $N_{\text{mod.}}$. For the numerical implementation we choose $\Omega = 1$ GeV and $N_{\text{mod.}}=1.1$.  

We compare the analytic predictions against \textsc{Pythia} simulation in \fig{hadronization}. We denote the end of SDNP region with a vertical dashed line which is located at 
\begin{equation}
    \mgr = \frac{\Lambda_{\text{NP}}^2}{\zcut}\,.
\end{equation}
The hadronic and partonic spectra of the simulation are normalized to the tail of the distributions ($\mgr \sim Q^2$) where non-perturbative effects are anticipated to have small effect. Then the NLL prediction is normalized to the region $e_{\min} < \mgr < e_{\max}$ as described above. The normalized NLL distribution is then convolved to the shape function as in \eq{shape_convolution}. We find that that normalizing the distribution to the tail is necessary for the $N_{\text{mod.}}$ to be independent of the grooming parameter, $\zcut$. Within the SDNP region the agreement with the simulation is within 5 to 10\%, and since in this region the shape of the cross-section is independent of the incoming hadronic matrix element it offers a unique opportunity for detail hadronization studies. Significant deviation is found for $\mgr \lesssim 1\; \text{GeV}^2$ where hadron mass effects are anticipated to play a significant role. 

%%%%%%%%%%%%%%%%%%%%%%%%%%%%%%%%%%%%%%%%%%%%%%%%%%%%%%%%%%%%%%%%%%%%%%%%%%%%%%%%%%
\section{An event based Winner-Take-All axis}\label{sec:WTA}

In the spirit  of event-based definition of grooming, in this section we extent the discussion to the recoil-free Winner-Take-All (WTA) axis. We will not discuss in detail applications of such an axis-finding procedure but we point to some opportunities in the context of TMD physics and 3D-imaging of the nucleon and nuclei. 

The WTA axis~\cite{Bertolini:2013iqa,Salam:WTAUnpublished} was introduced in the context of jets as an alternative recombination scheme which results to a recoil free jet axis. In contrast to the standard jet axis (SJA), which is defined by the sum of all particles within the jet, WTA axis follows the direction of the most energetic branch during the jet-clustering procedure. The resulting axis is then insensitive to soft radiation within the jet and thus free from underlying event contamination. 

In DIS and lepton colliders the WTA axis was proposed~\cite{Neill:2016vbi,Gutierrez-Reyes:2018qez,Gutierrez-Reyes:2019vbx} as a probe to quark TMD-PDFs and TMD-FFs. In this case the advantage of using this particular axis instead of the SJA is that in contrast to SJA, a recoil-free axis is insensitive to non-global logarithms (NGLs) which they could induce large systematic uncertainties.  It was also discussed in refs.~\cite{Gutierrez-Reyes:2019msa,Gutierrez-Reyes:2018qez} that a recoil free axis is in general less sensitive to hadronization corrections compared to the SJA and thus permit for a cleaner extraction of initial and final state TMDs. In ref.~\cite{Chien:2020hzh} it was also discussed the insensitivity of the WTA based observables to tracks versus all-particle measurements.

Here we discuss the construction of a WTA axis using an event level clustering procedure. In practice our approach differs from the standard definition of WTA axis since here one does not construct any jets and rather it applies an axis-finding algorithm directly to the event. The algorithm reads as follows,  
\begin{enumerate}
    \item For every pair of particles $\{i,j\}$ we calculate the Centauro measure, 
    \begin{align}
       %\hspace*{1cm}  
       d_{ij} &= (\Delta \bar{\eta}_{ij})^2 +2 \bar{\eta}_i \bar{\eta}_j (1-\cos \Delta \phi_{ij})\;, 
    \end{align}
    \item We find the minimum of all  $d_{ij}$ and we merge the particles $i$ and $j$ into a new ``branch''. The momentum of the new branch is given by the magnitude of the sum of the momenta of particles/branches $i$ and $j$ but is aligned along the direction of the one with the highest energy fraction $z$, defined in \eq{mom-var}.
    \item Repeat until all particles in the event are merged together. 
\end{enumerate}
When all particles are merged together the direction of momenta of the final single branch is the WTA axis as we define it in this work. 

In the back-to-back limit the finding algorithm we propose here returns a similar axis as if one reconstructed the WTA-axis of the leading jet for $R\sim 1$ (where $R$ is the jet radius) using the conventional definition. However, in multiple-jet configurations or for $R \ll 1$  the jet based definition it returns one axis per jet while the event-based algorithm returns a unique axis.

In the Breit frame and in the back-to-back limit the angle between the WTA axis and the virtual photon is sensitive to the universal TMD-PDFs and the rapidity anomalous dimension.
Note that this measurement involves the back-to-back soft function, the same matrix element that appears in the conventional TMD processes. This ensures the universality of the TMDPDFs relevant for this process to the ones that appear in conventional SIDIS and Drell-Yan, in contrast to other proposed observables that involve lepton-jet correlations in the Laboratory frame and involve soft functions that are sensitive to the jet kinematics \cite{Liu:2018trl, Liu:2020dct}. 
The details for the theoretical treatment of the proposed observable is discussed in section 2.3 of  ref.~\cite{Gutierrez-Reyes:2019vbx} and the formalism is directly applicable here.

%%%%%%%%%%%%%%%%%%%%%%%%%%%%%%%%%%%%%%%%%%%%%%%%%%%%%%%%%%%%%%%%%%%%%%%%%%%%%%%%%%
\section{Summary and outlook}\label{sec:conclusions}
% Summary
In this paper we discussed a novel grooming procedure for deep inelastic scattering (DIS) events. The corresponding algorithm we propose here is an extension of modified MassDrop Tagger (mMDT). Some of the main differences between our proposal and mMDT are: (in our proposal) i) clustering is done using Centauro measure instead of the C/A algorithm, ii) the algorithm takes as an input the DIS event in the Breit frame rather than a jet, and iii) the grooming condition uses the light-cone component of the momenta rather than the energy or transverse momentum.  With these modifications, we can effectively remove radiation close to the direction of the proton which usually consists of beam remnants and initial state radiation. In practice, it also removes soft particles, isolating this way the energetic final state radiation which comes from the fragments of the struck quark in a leading order DIS process.\footnote{For recent developments on event-level groomed event shapes in hadronic colliders see~\cite{Baron:2020xoi}.}  We demonstrate the applicability of this grooming procedure by considering an event shape measurement. Particularly, we investigate the groomed invariant mass (GIM), $\mgr$, which is a measure of how well collimated are the  particles that pass grooming. In the back-to-back limit (i.e., $\mgr \ll Q^2$), this observable is (up to power-corrections) directly related to the groomed event shape 1-jettiness. We show that the GIM spectrum is insensitive to rapidity cutoffs in the Laboratory frame usually imposed  due to detector acceptance. We derive factorization theorems within SCET for the back-to-back limit considering two different hierarchies: region 1 where $1\gg\zcut \gg \mgr/Q^2$ and region 2 where $1\gg\zcut \sim \mgr/Q^2$. We use the factorized forms to evaluate the resummed cross-section at NLL and compare against the partonic shower of \textsc{Pythia} 8. We find excellent agreement within the perturbative regime. At NNLL we show that the theoretical uncertainty  for the normalized cross-section is about 10\%. Furthermore we discuss hadronization effects and we show that these effects (for $\mgr \zcut \lesssim \Lambda_{\text{QCD}}$) can be well described with a shape function convolved with the resummed spectrum.

% On other event shapes
Although in this work we only consider the measurement of GIM,  one may also consider a plethora of other event shape measurements (e.g., N-jettiness and angularities) or even novel event-level generalizations of jet substructure measurements. An interesting example is groomed jet radius $R_g =d_{ij}$, which corresponds to the geometric separation of the two final branches, $i$ and $j$, that stopped the pruning. In the case of jet grooming  this observable has been studied extensively both experimentally~\cite{Adam:2020kug} and theoretically~\cite{Cacciari:2008gn,Larkoski:2014wba,Kang:2019prh}. In DIS and using the proposed algorithm we can generalize this observable to event level grooming. 

% Discussion on eA
Groomed jet substructure observables such as the energy fraction sharing $z_g$ and the groomed radius $R_g$ are particularly sensitive to the parton shower evolution of the jet. They are therefore great observables to nuclear medium induced modifications to the evolution of jets. The CMS~\cite{Sirunyan:2017bsd}, ALICE~\cite{Acharya:2019djg}, and STAR~\cite{Kauder:2017cvz} collaborations have already measured the modification to the $z_g$ spectrum in heavy-ion coalitions, although, the interpretation of these modification remains a subject of debate~\cite{Mehtar-Tani:2016aco,Chien:2016led,Milhano:2017nzm,Chang:2017gkt}. The future Electron-Ion-Collider (EIC) will offer a unique opportunity for the study of nuclear effects in $eA$ collisions with unprecedented accuracy. These observables in the context of groomed jets have been discussed in ref.~\cite{Arratia:2019vju}. We propose the observables, $z_g$ and $R_g$, adapted to event grooming in DIS (proposed in this work) as a probe to cold nuclear matter effects at the EIC.

%TMDs
Beyond event shapes and jet substructure, the use of groomed jets as probes to TMDs is an other interesting concept that has been gaining significant attention recently~\cite{Makris:2017arq,Makris:2018npl,Gutierrez-Reyes:2019msa}, particularly in the context of EIC. Generalization of this observables to event level grooming it is also possible. One can consider the angle between the virtual photon and the groomed axis (as a probe to TMDPDFs) or the transverse momentum of an identified hadron w.r.t. the groomed axis (as a probe to TMDFFs and TMD evolution).\footnote{Groomed axis in the context of event grooming in DIS is defined by the axis pointing along the direction of the total three-momenta of given by the sum of all particles that passed grooming.}

%WTA
Finally we give an event based definition of the Winner-Take-All (WTA) axis using the Centauro measure to cluster the whole event.  We emphasize that initial state universal TMDs (such as unpolarized  TMDPDFs and quark Siver's function) can be accessed by measuring the photon transverse momentum w.r.t. the WTA axis, similarly as discussed above for the groomed jet axis. The theoretical framework for such a measurement has been discussed in refs.~\cite{Gutierrez-Reyes:2018qez,Gutierrez-Reyes:2019vbx}. Furthermore, three dimensional fragmentation can also be accessed through  measurements of the transverse momentum of an identified hadron w.r.t. the WTA axis~\cite{Neill:2016vbi}.
\vspace{0.3 cm}

\begin{acknowledgments}
We thank Miguel Arratia and Duff Neill for feedback on the manuscript. Y.M. is supported by the European Union’s Horizon 2020 research and innovation programme under the Marie Skłodowska-Curie grant agreement No. 754496 - FELLINI.
\end{acknowledgments}

%%%%%%%%%%%%%%%%%%%%%%%%%%%%%%%%%%%%%%%%%%%%%%%%%%%%%%%%%%%%%%%%%%%%%%%%%%%%%%%%%%
\onecolumngrid
\appendix*
\section{Fixed order ingredients}
Here we collect from the literature all the fixed order ingredients we used for the construction of the resummed cross-section. It is organized in five subsections: hard, bean, jet,  collinear-soft, and ultra-soft functions. The soft function which is scaleless at this order is not discussed. At the end of the appendix we also give the leading order full-QCD cross section for the groomed invariant mass. 
%%%%%%%%%%%%%%%%%%%%%%%%%%%%%%%%%%%%%%%%%%%%%%%%%%%%%%%%%%%%%%%%%%%%%%%%%%%%%%%%%%
\subsection{Hard functions}
The relevant DIS hard function given in terms of the photon's virtuality is
\begin{equation}
    H(Q,y, \mu )  = \sigma_0(Q,y) \lbc 1 + a_s \lb - 2 \Gcusp^0 \ln^2 \lp \frac{\mu}{Q} \rp + \gamma^0_H \ln \lp \frac{\mu}{Q} \rp  + C_F \lp -16 +\frac{ \pi^2}{3}\rp \rb  + \mathcal{O}(\alpha_s^2)\rbc\;,
\end{equation}
where 
\begin{equation}
    \sigma_0(Q,y) = 2\pi \alpha_{e}^2 \frac{1+(1-y)^2}{Q^4}
\end{equation}
The hard function satisfies the following RGE:
\begin{equation}
    \frac{d}{d\ln\mu} H(Q, y,\mu ) = \gamma_H(Q,\mu) H(Q,y, \mu )\;,
\end{equation}
with $\gamma_H$ the hard function anomalous dimension. following the standard notation we may decompose the anomalous dimension into a term proportional to the cusp anomalous dimension, $\Gamma_{\text{cusp}}$, and a non-cusp terms and they have the following, 
\begin{equation}
    \gamma_{H}(Q,\mu) = - 4 \Gcusp[a_s] \ln\lp\frac{\mu}{Q} \rp +  \gamma_{H}[a_s]\;,
\end{equation}
where for the expansion in the strong coupling of all anomalous dimensions we are using the following notation
\begin{equation}
     \gamma_{G}[a_s] = \sum_{n = 0}^{\infty} a_s^{n+1} \gamma^{n}_G\;,
\end{equation}
where the subscript $G$ is a placeholder index for a generic function. For up to the NNLL cross-section we need the following coefficients,
\begin{align}
\Gcusp^0 &= 4 C_F\,,\nl
\Gcusp^1 &= 4 C_F \lb C_A\,\lp \frac{67}{9} -\frac{\pi^2}{3} \rp  -
   \frac{20}{9}\,T_F\, n_f \rb \,,\nl
\Gcusp^2 &= 4 C_F \lb
C_A^2 \lp\frac{245}{6} -\frac{134 \pi^2}{27} + \frac{11 \pi ^4}{45}
  + \frac{22 \zeta_3}{3}\rp + C_A\, T_F\,n_f \lp- \frac{418}{27} + \frac{40 \pi^2}{27}  - \frac{56 \zeta_3}{3} \rp \nl
  &   + C_F\, T_F\,n_f \lp- \frac{55}{3} + 16 \zeta_3 \rp 
  - \frac{16}{27}\,T_F^2\, n_f^2 \rb \;,
\end{align}
and 
\begin{align}
\gamma_{H}^0 &= -12 C_F \,,\nl
\gamma_{H}^{1}
&= C_F \lb C_A \lp-\frac{164}{9} +104 \zeta_3\rp
+ C_F (-6 +8 \pi^2 - 96 \zeta_3) + \beta_0 \lp-\frac{130}{9} -2 \pi^2 \rp \rb\;.
\end{align}
%%%%%%%%%%%%%%%%%%%%%%%%%%%%%%%%%%%%%%%%%%%%%%%%%%%%%%%%%%%%%%%%%%%%%%%%%%%%%%%%%%
\subsection{Groomed beam function}
At NLO there is only one final state gluon and therefore no clustering effects enter at this order. The beam function is then given by the \emph{cumulant} beam-thrust beam function with the replacement $t_{\text{cut}} \to Q^2 \zcut$. This result can be found in ref.~\cite{Stewart:2009yx}
\begin{equation}
   \mathcal{B}_{q/P} (x,Q^2\zcut,\mu) = \sum_i \int_x^{1} \frac{dz}{z} \,\mathcal{I}_{qi}\lp \frac{x}{z},Q^2\zcut,\mu \rp f_{i/P}(z,\mu)\;,
\end{equation}
where the short distance matching coefficients are 
\begin{equation}
    \mathcal{I}_{qi}(x,Q^2\zcut,\mu)= \delta_{qi} \delta(1-x)  \\[5pt] 
    + a_s \lbc  \frac12 \delta_{qi} \delta(1-x)  \lb \Gcusp^0 \LB + \gamma_B^0 \rp  \rb \LB  -\LB P_{qi}(x) + \mathcal{I}_{qi}^{(1)}(x)  
     \rbc\;.
\end{equation}

The one-loop QCD splitting kernels are,
\begin{align}\label{eq:QCD-splitting}
    P_{qq}(x) &= 2C_F\lp \frac{1+x^2}{1-x}\rp_+ \;, & P_{qg}(x) &=2T_F(1-2x(1-x))\;,
\end{align}
and the non-singular terms are, 
\begin{align}
    \mathcal{I}^{(1)}_{qq}(x) &= 2C_F \lb (1+x^2)\lp \frac{\ln(1-x)}{1-x}\rp_{+} -\frac{\pi^2}{6}\delta(1-x) +1-x -\frac{1+x^2}{1-x} \ln(x)     \rb\;,\nl
    \mathcal{I}^{(1)}_{qq}(x) &= 2T_F + P_{qg}(x) \lb \ln\lp \frac{1-x}{x} \rp -1  \rb\;.
\end{align}
The beam function satisfies the following RGE,
\begin{equation}
    \frac{d}{d\ln \mu} \mathcal{B}_{q/P}(x,Q^2 \zcut,\mu) = \gamma_{\mathcal{B}}(Q^2 \zcut,\mu) \mathcal{B}_{q/P}(x,Q^2 \zcut,\mu)\;,
\end{equation}
where
\begin{equation}
    \gamma_{\mathcal{B}}(Q^2 \zcut,\mu) = 2 \Gcusp[a_s] \LB +\gamma_{\mathcal{B}}[a_s]\;,
\end{equation}
with
\begin{align}
    \gamma_{\mathcal{B}}^{0} & = 6 C_F \;,\nl
\end{align}
As discussed earlier the two-loop beam anomalous dimension, $\gamma_{\mathcal{B}}^1$, is unknown and requires a non trivial computation considering clustering effect. However the combination, $\gamma_{\mathcal{B}}^1 + \gamma_S^1$, we can obtain from consistency relations, 
\begin{align}
     \gamma_{\mathcal{B}}^{1} + \gamma_S^1  &= C_F \lb
  C_A \lp-7.73 - 80 \zeta_3\rp 
+ C_F (20 - 4 \pi^2 + 48 \zeta_3) + \beta_0 \lp 18.71 + \frac{2\pi^2}{3} \rp  \rb\;.
\end{align}
Note for resumming all logarithms $\ln(\zcut)$ at NNLL requires the knowledge one of $\gamma_{\mathcal{B}}^1$ or $\gamma_{S}^1$ separately.
%%%%%%%%%%%%%%%%%%%%%%%%%%%%%%%%%%%%%%%%%%%%%%%%%%%%%%%%%%%%%%%%%%%%%%%%%%%%%%%%%%
\subsection{Jet function}
The same jet function appears in the factorization theorem for both regions 1 and 2. In fact the jet function that appears in these factorization theorems is the same as the jet function that appears the jet-mass measurements~\cite{Bauer:2003pi} and has been calculate up to three-loop accuracy. Here we collect the one-loop fixed order results and the two-loop anomalous dimension.  In terms of the  variable $u$ Laplace conjugate of the invariant mass (in literature usually denoted with $s$) we have,
\begin{equation}
    J(u, \mu^2) = 1 + a_s \lb \frac{1}{2} \Gcusp^0 \ln^2 (u\,\mu^2) + \frac{1}{2} \gamma^0_J \ln (u\,\mu^2)
    + C_F \lp 7 -2\frac{\pi^2}{3}\rp \rb\;,
\end{equation}
and satisfies the following RGE:
\begin{equation}
    \frac{d}{d\ln\mu} J(u, \mu^2) = \gamma_J (u, \mu^2) J(u, \mu^2)\;,
\end{equation}
where
\begin{equation}
    \gamma_{J}(u, \mu^2) =  - 2 \Gcusp[a_s] \ln (u\,\mu^2) +  \gamma_{J}[a_s]\;,
\end{equation}
with
\begin{align}
\gamma_{J}^{0} &= 6 C_F \,,\nl
\gamma_{J}^{1}
&= C_F \lb
  C_A \lp\frac{146}{9} - 80 \zeta_3\rp 
+ C_F (3 - 4 \pi^2 + 48 \zeta_3)  + \beta_0 \lp \frac{121}{9} + \frac{2\pi^2}{3} \rp  \rb\;.
\end{align}
%%%%%%%%%%%%%%%%%%%%%%%%%%%%%%%%%%%%%%%%%%%%%%%%%%%%%%%%%%%%%%%%%%%%%%%%%%%%%%%%%%
\subsection{Collinear-soft function}
The collinear-soft function that appears in our analysis is the same function that also appears in the groomed jet-mass measurements for SoftDrop grooming for $\beta =0$. This is true since in the collinear limit the Centauro measure and the C/A clustering measure are converge. The one-loop result for the this collinear-soft function is given in eqs. (E.4) and (E.5) of ref.~\cite{Frye:2016aiz} (for our case $\alpha =2 $ and $\beta =0$), 
\begin{equation}
    \Csoft (m_{cs}^2\zcut,\mu ) = \frac{1}{N_C} \text{tr} \Langle \text{T}[Y_{n}^{\dag} W_t] \delta\lp m_{cs}^2 - \Theta_{\text{gr.}} \frac{Q^2 \hat{e}_{2}^{(2)}}{4} \rp \bar {\text{T}}[W_t^{\dag} Y_n] \Rangle = \frac{Q^2}{4} S_{C}(\zcut e_{2}^{(2)})\;,
\end{equation}
where $\Theta_{\text{gr.}}$ is a particle level selection function for particles that pass grooming. Taking the Laplace transform of the one-loop result with respect to $m_{cs}^2$ we have,
\begin{equation}
   \Csoft (u,\zcut,\mu ) = 1 -  a_s \frac{\Gcusp^0}{2} 
    \ln^2 \lp\frac{\mu^2 u}{ \zcut} \rp \;.
\end{equation}
This is the collinear-soft function at one loop for the groomed invariant mass measurement relevant to region 1. It satisfies the following RG, 
\begin{align}
    \frac{d}{d\ln \mu} \Csoft (u,\zcut,\mu ) &= \gamma_{\Csoft}(u,\zcut,\mu) \Csoft(u,\zcut,\mu )\,, 
\end{align}
where 
\begin{equation}
    \gamma_{\Csoft}(u,\zcut,\mu) = - 2 \Gcusp \ln \lp\frac{\mu^2 u}{ \zcut} \rp + \gamma_{\Csoft}[a_s] \;,
\end{equation}
with
\begin{align}
    \gamma_{\Csoft}^0 &= 0\;,\nl
    \gamma_{\Csoft}^1 &= C_F \lb
  C_A \lp9.73 + 56 \zeta_3\rp 
- C_F 17.00  + \beta_0 \lp - 17.71 + \frac{2\pi^2}{3} \rp  \rb\;.
\end{align}
%%%%%%%%%%%%%%%%%%%%%%%%%%%%%%%%%%%%%%%%%%%%%%%%%%%%%%%%%%%%%%%%%%%%%%%%%%%%%%%%%%
\subsection{Ultra-soft function}

\begin{figure}[t!]
  \centerline{\includegraphics[width =  0.8 \textwidth]{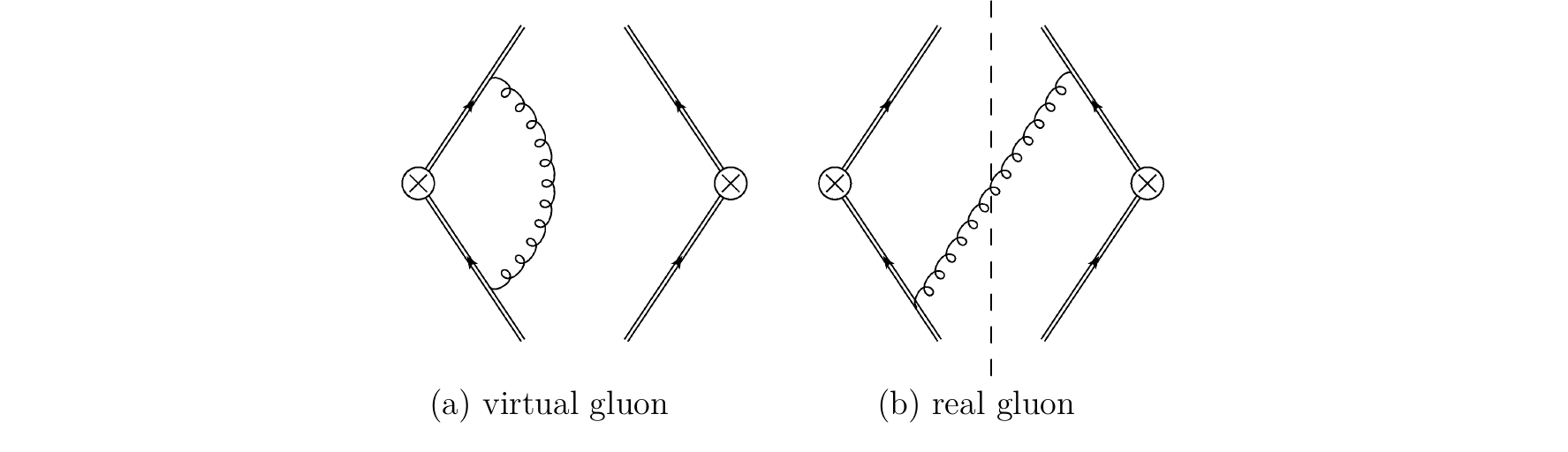}}
  \caption{~\label{fig:usoft-diagrams} Diagrams contributing at next-to-leading order in the perturbative calculation of the u-soft function. Mirror diagrams not shown in this figure.}
\end{figure}

We give the operator definition of the collinear ultra-soft function, $\Usoft$, That appears in the factorization theorem for region 2. We then continue with the one loop calculation and show that at this order gives the same result as the collinear-soft function. The operator definition of ultra-soft function is written in terms of  the SCET$_{\text{I}}$ ultra-soft Wilson-lines (hence the name ultra-soft function) along the $n$ and $\bn$ directions,
\begin{equation}
    \Usoft (m_{us}^2,Q,\zcut,\mu ) = \frac{1}{Q\,N_C} \text{tr} \Langle \text{T}[Y_{n}^{\dag} Y_{\bn}] \delta\lp \frac{m_{us}^2}{Q} - \hat{\mathcal{M}}_{\text{gr.}}^- \rp \bar {\text{T}}[Y_{\bn}^{\dag} Y_n] \Rangle \;,
\end{equation}
where $\hat{\mathcal{M}}_{\text{gr.}}^- $ its is the operator that returns the $\bn \cdot p$ component of the total momenta of the particles that pass grooming. We can cont give a close form for this operator at all orders in perturbation theory since it depends on the number of final state particles, but it can be defined at the Feynman diagram level. At leading order there are no virtual or real gluons and all Wilson-lines are set to the identity operator and thus we get,
\begin{equation}
    \Usoft (m_{us}^2,Q,\zcut,\mu )  = \delta(m^2_{us})+ \mathcal{O}(\alpha_s)\,,
\end{equation}
At next-to-leading order we need to consider the diagrams shown in \fig{usoft-diagrams}. Diagram (a)  corresponds to the virtual gluon exchange between the Wilson-lines in the $n$ and $\bn$ directions. In pure dimensional regularization this yields a scaleless integral and thus vanishes in our scheme. Diagram (b) involves one real gluon exchange. The contribution of that gluon with momentum, $k^{\mu}$, to the invariant mass can be obtained by considering the sum of all radiation that pass grooming, 
\begin{equation}
    \mgr = \lp \sum_{i} p_{i}^{\mu} \rp = (p_{\bn}^{\mu} + k^{\mu})^2 = p_{\bn}^2 + Q k^{-}\;.
\end{equation}
Assuming that the gluon passes grooming there are two terms relevant to the invariant mass measurement: the contribution from the $\bn$-collinear radiation, and the interference term  $Q k^-$ which involves the virtual gluon. Therefore, the diagram (b) contribution to the one-loop u-soft function is given by
\begin{equation}
  d_{\text{(b)}} =  g^2 C_F \lp \frac{e^{\gamma_E}  \mu^2}{4\pi} \rp^{\epsilon} \int\frac{d^d k}{(2\pi)^{d-1}} \frac{1}{k^- k^+} \delta(k^2) \lb \delta(m^2_{us} - Q k^- ) \Theta(k^+ - Q \zcut ) + \delta(m^2_{us}  ) \Theta( Q \zcut -k^+ )  \rb \;.
\end{equation}
The first term in the square brackets corresponds to the case where the ultra-soft gluon passes grooming and the second term in the case where it fails. The second term also yield a scaleless integral and thus ignore in our calculation. Performing the trivial integration and multiplying with a factor if two (for the mirror diagram) we find that the u-soft and collinear-soft functions are identical at this order.

%%%%%%%%%%%%%%%%%%%%%%%%%%%%%%%%%%%%%%%%%%%%%%%%%%%%%%%%%%%%%%%%%%%%%%%%%%%%%%%%%%
\subsection{Fixed order QCD at $\bmat{\mathcal{O}(\alpha_s)}$ }

In this section we give the main aspects and result of the $\mathcal{O}(\alpha_s)$ calculation of the GIM cross section in QCD. For this task we will closely follow the derivation and notation of ref.~\cite{Kang:2014qba} which was suited for the calculation of 1-jettiness in DIS but also very adaptable to our calculation. The differential cross section can be written in terms of the hadronic tensor, 
\begin{equation}\label{eq:fullQCD}
    \frac{d\sigma}{dx dQ^2 d\mgr}=\frac{\alpha^2}{2 Q^4} \lp [1+(1-y)^2] W_G(x,Q^2,\mgr) +2x^2[6-6y+y^2]W_P(x,Q^2,\mgr)  \rp\,,
\end{equation}
where 
\begin{align}
    W_G (x,Q^2,\mgr)&= - g_{\mu \nu} W^{\mu \nu}(x,Q^2,\mgr) \,,&  W_P(x,Q^2,\mgr) &= \frac{P_\mu P_\nu}{Q^2} W^{\mu \nu} (x,Q^2,\mgr)\,,
\end{align}
which we refer to as the $G$ and $P$ projections of the hadronic tensor.  For groomed invariant mass measurement the hadronic tensor is written in terms of the QCD  quark current operator, $J^{\mu}(r)$, as follows:
\begin{equation}
    W^{\mu\nu}(x,Q^2,\mgr) = \int d^{4}r \exp(i q \cdot r) \langle P | {J^{\mu}}^{\dag}(r) \delta(\mgr - \hat{\mathcal{M}}_{\text{gr.}}) J^{\nu}(0) | P \rangle\,.
\end{equation}
The operator $\hat{\mathcal{M}}_{\text{gr.}}$ projects the value of the hadronic GIM for a given final state. Our goal in this section is to give explicit results for $W_G$ and $W_P$.  

The hadronic tensor can be matched onto the collinear parton distribution functions in an operator product expansion (OPE), 
\begin{equation}\label{eq:OPE}
    W_j(x,Q^2,\mgr) = \sum_{i\in \{q,\bar{q},g\}} \int_{x}^{1}\frac{d\xi}{\xi} f_{i/P}(\xi,\mu) \,w^{i}_{j} \lp \frac{x}{\xi},Q^2,\mgr,\mu\rp \times \lb 1+\mathcal{O}\lp \frac{\Lambda_{\text{QCD}}}{\sqrt{\mgr}} \rp \rb
\end{equation}
where $w^{i}_{j}$ are the perturbative matching coefficients calculable order-by-order in perturbation theory.  We now discuss the calculation of the $\mathcal{O}(\alpha_s)$ (leading order) $w^{i}_{j}$ functions.  We consider only finite values of the GIM (i.e., $\mgr > 0$) and thus drop any regulator dependence and only consider the real emissions. We therefore have two final state partons that contribute to the measurement. Their momenta (working in the Breit frame) are given by,
\begin{align}
    p_1^{\mu} & = Q(1-v)\frac{\bn^{\mu}}{2} + \frac{1-x}{x}Qv \frac{n^{\mu}}{2} - p_{\perp}^{\mu} \nl   
    p_2^{\mu} & = Qv\frac{\bn^{\mu}}{2} + \frac{1-x}{x}Q(1-v) \frac{n^{\mu}}{2} + p_{\perp}^{\mu}
\end{align}
and $p_{\perp}^{\mu}$ can be obtained from the on-sell condition $p_1^2 =p_2^2 = 0$. Thus for fixed values of $x\text{-}Q^2$ the only free parameter is $v$ defined as $v\equiv p_2^+/Q$. The grooming algorithm will first cluster these two particles and in the de-clustering process the grooming condition reads,
\begin{equation}
    \min[v,\,1-v] > \zcut\,.
\end{equation}
Therefore if this condition is satisfied both particles will pass grooming and contribute to the measurement,
\begin{equation}
    \mgr = (p_1+p_2)^2 = \frac{1-x}{x}Q^2\,.
\end{equation}
Otherwise, if the condition is not satisfied only one particle pass grooming yielding $\mgr = p_i^2 = 0$ and thus does not contribute to the finite GIM spectrum. 
We then have for the matching coefficients, 
\begin{align} \label{eq:matching-all}
    w^{i}_{\mu \nu} (x,Q^2,\mgr,\mu) \Big{|}_{\mathcal{O}(\alpha_s)} = -\frac{1}{16\pi} \delta \lp \mgr - \frac{1-x}{x}Q^2 \rp \int_{\zcut}^{1-\zcut} dv \mathcal{M}^{\text{real}}_{\nu}(v,Q^2,\mu) \lp\mathcal{M}^{\text{real}}_{\mu}(v,Q^2,\mu) \rp^{*}
\end{align}
with $\mathcal{M}^{\text{real}}_{\nu}$ the QCD amplitude from real emission in the process $\gamma^* i \to 12$. The projected matching coefficients that appear in \eq{OPE} are given by  
\begin{align}
    w^i_G (x,Q^2,\mgr,\mu)&= - g^{\mu \nu} w^i_{\mu \nu}(x,Q^2,\mgr,\mu) \,,&  w^i_P(x,Q^2,\mgr,\mu) &= \frac{P^\mu P^\nu}{Q^2} w^i_{\mu \nu} (x,Q^2,\mgr,\mu)\,,
\end{align}
the equivalent projections of the amplitude squared and the integration over $v$ can be done using the expressions in eqs.(B.8-B.10) of ref.~\cite{Kang:2014qba}. Finally the integration over $\xi$ in \eq{OPE} can be trivially performed using the measurement delta function in \eq{matching-all}. Therefore the  $G$ and $P$ projections of the hadronic tensor are 
\begin{equation}
    W_j(x,Q^2,\mgr) = \alpha_s(\mu) \;\frac{2x}{\xi Q^2} \; \Theta(x < \xi < 1)\sum_{i \in \{q,\bar{q} ,g \}} Q^{2}_i\; f_{i/p}(\xi,\mu) \, h^i_j\lp\frac{x}{\xi},\zcut \rp
\end{equation}
with $\xi = x (1+\mgr/Q^2)$ and 
\begin{align}
    h^{q}_{G}(x,\zcut)  &= C_F\; \frac{1}{2(1-x)} \lp (1-4x)(1-2\zcut)+2(1+x^2) \ln \frac{1-\zcut}{\zcut}  \rp \nl
    h^q_{P}(x,\zcut)  &= C_F \; \frac{1}{4 x} (1-2 \zcut)  \nl
    h^g_{G}(x,\zcut)  &= T_F \;\lp -2 (1-2\zcut) +4 (1-2x + 2x^2) \tanh^{-1} (1-2\zcut) \rp \nl
    h^g_{P}(x,\zcut)  &= T_F \; \frac{1-x}{x} (1-2\zcut) 
\end{align}
and $h^{\bar{q}}_{G/P}= h^{q}_{G/P}$. For quarks and antiquarks the charge $Q_i^2$ is simply the electric charge of that parton and for gluons we have, 
\begin{equation}
    Q_g^2 \equiv   \sum_{f} Q_f^{2} 
\end{equation}
where the sum runs over the quark flavors only.

%%%%%%%%%%%%%%%%%%%%%%%%%%%%%%%%%%%%%%%%%%%%%%%%%%%%%%%%%%%%%%%%%%%%%%%%%%%%%%%%%%
\twocolumngrid

\normalbaselines 
\bibliography{main}
%%%%%%%%%%%%%%%%%%%%%%%%%%%%%%%%%%%%%%%%%%%%%%%%%%%%%%%%%%%%%%%%%%%%%%%%%%%%%%%%%%

\end{document}